\documentclass[conference]{IEEEtran}
\usepackage{cite}
\usepackage{amsmath,amssymb,amsfonts}
\usepackage{algorithmic}
\usepackage{graphicx}
\usepackage{textcomp}
\usepackage[hyphens]{url}
\usepackage{fancyhdr}
\usepackage[bookmarks=true,breaklinks=true,letterpaper=true,colorlinks,citecolor=blue,linkcolor=blue,urlcolor=blue]{hyperref}

\usepackage{mathptmx} % This is Times font
\usepackage{cite}
\usepackage{wrapfig}
\usepackage{amsmath,amssymb,amsfonts}
\usepackage{booktabs}
\usepackage{caption}
\usepackage{graphicx}
\usepackage{subcaption}
\usepackage{listings}
\usepackage[table]{xcolor}
\usepackage{tcolorbox}
\usepackage{textcomp}
\usepackage{boldline}
\usepackage{float}
\usepackage{xspace}
\usepackage{booktabs, makecell, multirow, tabularx}
\usepackage{todonotes} % for comments4
\usepackage{wrapfig,lipsum}
\usepackage{verbatim}
\usepackage{graphicx}
\usepackage{listings}
\usepackage{fancyhdr}
\usepackage[normalem]{ulem}
\usepackage[hyphens]{url}

\usepackage{tikz}                   
\usetikzlibrary{shadows}
\usepackage{fancybox} 
\usepackage{tcolorbox}
\usepackage{array}
\tcbuselibrary{skins}
\definecolor{royalblue}{RGB}{65,105,225}

\newcommand{\fname}{{\em SecScale}\xspace}

\newcommand*\circlenew[1]{\tikz[baseline=(char.base)]{
            \node[shape=circle,draw,inner sep=0.3pt] (char) {#1};}}  
\definecolor{royalblue}{RGB}{65,105,225}
\fancypagestyle{firstpage}{
  \fancyhf{}
  
  \fancyfoot[C]{\thepage}
}

\definecolor{royalblue}{RGB}{65,105,225}
\title{\fname: A Scalable and Secure Trusted Execution Environment for Servers} 
\author{\IEEEauthorblockN{Ani Sunny}
\IEEEauthorblockA{Computer Science and Engineering\\
Indian Institute of Technology Delhi\\
New Delhi, India\\
csz228276@cse.iitd.ac.in}
\and
\IEEEauthorblockN{Nivedita Shrivastava}
\IEEEauthorblockA{Electrical Engineering\\
Indian Institute of Technology Delhi\\
New Delhi, India\\
nivedita.shrivastava@ee.iitd.ac.in}
\and
\IEEEauthorblockN{Smruti R. Sarangi}
\IEEEauthorblockA{Computer Science and Engineering\\
Indian Institute of Technology Delhi\\
New Delhi, India\\
srsarangi@cse.iitd.ac.in}}
\begin{document}
\maketitle
\thispagestyle{firstpage}
\pagestyle{plain}

%%%%%% -- PAPER CONTENT STARTS-- %%%%%%%%

%\and
%\IEEEauthorblockN{Nivedita Shrivasthava}
%\IEEEauthorblockA{Electrical and Electronics Engineering\\
%Indian Institute of Technology Delhi\\
%India\\
%Email: eez198358@ee.iitd.ac.in}
%\and
%\IEEEauthorblockN{S. R. Sarangi}
%\IEEEauthorblockA{Computer Science and Electrical Engineering\\
%Indian Institute of Technology Delhi\\
%India\\
%Email: srsarangi@cse.iitd.ac.in}}

\begin{abstract}
Trusted execution environments (TEEs) are an integral part of modern secure processors. They ensure that their application and code pages are confidential, tamper-proof and immune to diverse types of attacks. In 2021, Intel suddently announced its plans to deprecate its most trustworthy enclave, SGX, on its $11^{th}$ and $12^{th}$ generation processors. The reasons stemmed from the fact that it was difficult to scale the enclaves (sandboxes) beyond 256 MB -- the hardware overheads outweighed the benefits. Competing solutions by Intel and other vendors are much more scalable, but do not provide many key security guarantees that SGX used to provide notably replay attack protection. In the last three years, no proposal from industry or academia has been able to provide both scalability (with a modest slowdown) as well as replay-protection on generic hardware (to the best of our knowledge). We solve this problem by proposing \fname that uses some new ideas centered around speculative execution (read-first, verify-later), creating a forest of MACs (instead of a tree of counters) and providing complete memory encryption (no generic unsecure regions). We show that we are 10\% faster than the nearest competing alternative.
\end{abstract}

\section{Introduction}
The number attacks on remotely executing software
in both public and private clouds is on the rise\cite{cloud_attack}.
Along with software-based attacks, a large number of physical attacks
such as cold boot attacks and bus snooping are also being mounted~\cite{coldboot}.
According to an IBM report\cite{ibm}, the
cost of a data breach in 2023 was $\$$4.5 million and
{\em 82$\%$} of the data breaches involved data that was stored in the {\em cloud}. 

To secure data and computation in any such remote framework, we need to use a combination of
encryption, message authentication codes (MACs), and digital signatures, respectively,
for ensuring the following four ACIF properties: authenticity(A), confidentiality(C), integrity(I)  and freshness(F). Authenticity refers to the fact that the data was indeed written by the
server's CPU; confidentiality uses encryption to prevent snooping; integrity prevents tampering (using
hashes and keyed hashes(MACs) and freshness ensures that data that was valid in the past
is not being \emph{replayed}.
Table~\ref{table:enclave} shows a list of all the major commercially available TEEs including SGX.

\begin{scriptsize}
\begin{table*}[!htb]
\footnotesize
    \centering
   \setlength\tabcolsep{1.5pt} 
\begin{tcolorbox}[enhanced, width=.67\textwidth, boxsep=0pt, left=0pt,right=0pt,top=0pt,bottom=0pt,
    colback=white, colframe=black, arc=0mm, boxrule=0pt,
    drop shadow={shadow xshift=100mm, shadow yshift=100mm}]
 \rowcolors{2}{royalblue!20}{royalblue!5}
    \begin{tabular}{|l|c|c|c|c|c|c|c|c|}
    \hline
     \rowcolor{gray!10}

    \hline
    \rowcolor{gray!10}
         \textbf{System} & \textbf{Arch.} & \textbf{Intro.} &\textbf{Memory}  &\textbf{Integrity} &\textbf{Freshness} & \textbf{Unrestricted} & \textbf{Scalable} &  \textbf{Isolation} \\ %&\textbf{Remarks} \\
          \rowcolor{gray!10}
          \textbf{} & \textbf{} & \textbf{Year}&\textbf{Encr.}  &\textbf{} &\textbf{} & \textbf{Encl. Count} & &  \textbf{Granularity} \\ %&\textbf{} \\
         \hline           
                      
         \hline       
         ARM-TrustZone\cite{trustzone} & ARM & 2004 & \textcolor{red}{$\times$} & \textcolor{teal}{$\checkmark$} &  \textcolor{red}{$\times$} & \textcolor{teal}{$\checkmark$} & \textcolor{teal}{$\checkmark$} & World \\ %& Conf. and int. via isolation    \\
         
        \hline
         SGX-Client\cite{sgx} & Intel & 2015 & \textcolor{teal}{$\checkmark$} & \textcolor{teal}{$\checkmark$} &  \textcolor{teal}{$\checkmark$} & \textcolor{teal}{$\checkmark$} &  \textcolor{red}{$\times$} & Process \\ %& Ensures ACIF guarantees  \\
         
        \hline       
         SGX-Server\cite{sgx1} & Intel & 2016 & \textcolor{teal}{$\checkmark$} & \textcolor{teal}{$\checkmark$} &  \textcolor{red}{$\times$} & \textcolor{teal}{$\checkmark$} & \textcolor{teal}{$\checkmark$} & Process \\ %& Susceptible to replay attacks\\
          
        \hline       
         SEV\cite{sev} & AMD & 2016 & \textcolor{teal}{$\checkmark$} & \textcolor{red}{$\times$} &  \textcolor{red}{$\times$} & \textcolor{red}{$\times$} & \textcolor{teal}{$\checkmark$} & VM \\ %& Only data in-use protected   \\
   
          \hline       
         SEV-ES\cite{sev-es} & AMD & 2017 & \textcolor{teal}{$\checkmark$} & \textcolor{red}{$\times$} &  \textcolor{red}{$\times$} & \textcolor{red}{$\times$} & \textcolor{teal}{$\checkmark$} & VM \\ %& Only data in-use protected  \\
         
         \hline       
         SEV-SNP\cite{snp} & AMD & 2020 & \textcolor{teal}{$\checkmark$} & \textcolor{teal}{$\checkmark$} &  \textcolor{red}{$\times$} & \textcolor{red}{$\times$} & \textcolor{teal}{$\checkmark$} & VM \\ %&  Only data in-use protected   \\

         \hline     
         ARM-CCA\cite{arm_cca} & ARM & 2021 & \textcolor{red}{$\times$} & \textcolor{teal}{$\checkmark$} &  \textcolor{red}{$\times$} & \textcolor{teal}{$\checkmark$} & \textcolor{teal}{$\checkmark$} & Realm \\ %&  Conf. and int. via isolation  \\

         \hline       
         TDX\cite{tdx1} & Intel & 2023 & \textcolor{teal}{$\checkmark$} & \textcolor{teal}{$\checkmark$} &  \textcolor{red}{$\times$} & \textcolor{red}{$\times$} & \textcolor{teal}{$\checkmark$} & TD \\ %& SGX assisted remote auth. \\

         \hline     
         \textbf{\fname} & Intel & 2024 & \textcolor{teal}{$\checkmark$} & \textcolor{teal}{$\checkmark$} &  \textcolor{teal}{$\checkmark$} & \textcolor{teal}{$\checkmark$} & \textcolor{teal}{$\checkmark$} & Process \\ %&  Ensures ACIF guarantees  \\
    \hline
    \hline

    \end{tabular}
    \end{tcolorbox}
    \caption{Commercially available TEEs.
    Notes: \textbf{TD} refers to a virtualization-based trust domain, \textbf{Realm} and \textbf{World} are equivalent to VMs; Conf. refers to confidentiality, int. refers to integrity and auth. refers to authenticity.}
    \label{table:enclave}
\end{table*}
\end{scriptsize}

Among all the commercially available TEEs listed in Table~\ref{table:enclave}, now-deprecated 
Intel SGX\cite{sgx} (Software Guard Extensions)
is the only one that provides all four ACIF guarantees in HW (referred to as SGX-Client).
Third-party software on SGX-Client used to run in a HW-managed {\em enclave}
{\em securely} in spite of a potentially malicious OS or hypervisor.
However, this {\em robust} protection came at a heavy price.
The performance overheads limited the enclave size to 128-256 MB~\cite{benchsgx}.
As a result, Intel decided to deprecate SGX-Client in its $11^{th}$ and 
$12^{th}$ generation processors and supplanted it with SGX-Server\footnote{Both
SGX-Client and SGX-Server are terms that we introduce in this paper for the ease
of explanation.}. SGX-Server 
adopts a different mode of memory encryption and eliminates time-hungry integrity (Merkle)
trees altogether. It
scales to 512 GB; however, this scalability comes at the cost of security -- it is
possible to mount replay attacks~\cite{benchsgx,bl1}. 

This has sadly
impacted different industries and products quite adversely. For example,
ultra HD Blu-ray disks require the support of SGX's digital rights management (DRM) service\cite{bl6}. They can no longer be played on new Intel processors that only support SGX-Server\cite{bl5}.
There are similar issues with DRM-protected PC games\cite{bl9} and secure 4K video streaming apps
\cite{bl7}.
SGX-Client allowed these apps to run in an enclave and consequently guarantee that the viewer
wasn't able to steal video content~\cite{bl8}.
References~\cite{bl2,f10} contain a lot of examples of replay attacks in distributed
systems and software such as Ethereum and Bitcoin (attacks SGX-Client could prevent).
%It also turns out that replay attacks can used to amplify the power of other side-channel attacks \cite{bl3}. 
%FIXME: I am not very sure about this. You need to explain. 

There are two strands of contemporary work.
The first has been adopted by commercial silicon vendors who provide
security solutions primarily for VMs (virtual machines)~\cite{snp,tdx,arm_cca}.
The assumption is that
the entire guest VM is trustworthy including its software stack\cite{sev-tdx}. 
In the second strand of work, proposed in academia, two proposals stand out.
{\em Dynamic Fault History-Based Preloading (DFP)}\cite{dfp} implements a prefetching
based mechanism to improve performance and support larger enclaves.
Whereas,  
{\em Penglai}\cite{penglai} is a bespoke RISC-V system that relies on caching parts of the integrity
verification tree (Merkle tree) in a separate physical memory that has strict access protections.
The miss penalty is sadly quite high. Additionally, given that there is a large unrestricted unsecure
memory in the system, managing the page tables requires complex mechanisms.

The insights in our work \fname are as follows.
In any TEE, a security breach is a {\em catastrophic event} -- 
the entire system needs to shut down. Hence, we have the liberty to {\em speculate}:
read first and verify later. 
Second, existing work in this space creates a Merkle tree of counters, whereas we create a
MAC forest, which is paradigmatically quite different. Finally, we avoid the pitfalls of creating
one large SGX-like secure memory or splitting the physical address space into generic secure
and unsecure regions. The former approach is not scalable, which is why SGX-Client was
deprecated in the first place. We shall also show the same in our experiments. The latter approach 
adopted in conventional work
envisions having two parts of an executing application: secure part and an unsecure part. The secure
part has severe restrictions and the unsecure part functions as a regular program. This is a good strategy
when the size of the secure memory is limited. However, we propose fully encrypted memory (scales till 512 GB),
where the entire
memory is secure barring a few pages that are reserved for inter-process data transfer. Hence, there is no
need to split an application in this manner (also complicates the software design). To ensure inter-process
isolation, instead of relying on the OS to update page tables correctly, we rely on encryption with enclave
keys. We also use a more pragmatic threat model where we assume that the attacker can observe and modify any
location at will.

The specific \textbf{contributions} in this work are as follows:
\circlenew{1} Design of a scalable integrity protection mechanism, where we increase the granularity of the MAC computation from the block-level to the page-level and devise an efficient and scalable MAC forest based integrity protection mechanism.
\circlenew{2} Design of an efficient scheme for enclave page fault management that implements speculative execution and decreases the latency of the critical path.
\circlenew{3} An efficient encryption mechanism to secure the confidentiality of the entire memory.
\circlenew{4} A detailed performance and scalability analysis of
\fname, which shows a $10\%$ improvement over the nearest competing work (\textit{Penglai}). 

$\S$\ref{sec:back} introduces the necessary background.  $\S$\ref{sec:threat} outlines the
threat model, $\S$\ref{sec:char} characterizes the benchmarks and 
related work, $\S$\ref{sec:design}
presents the proposed design, $\S$\ref{sec:eval} shows 
a detailed performance analysis, $\S$\ref{sec:RW} presents the related work, and
we finally conclude in
$\S$\ref{sec:concl}. 
\vspace{-3mm}

\section{Background of Intel SGX}
\label{sec:back}

Intel Software Guard Extensions (SGX~\cite{sgx}) integrates hardware extensions to 
the x86 instruction set.
There are special instructions to create protected execution environments known as {\em enclaves}.
The enclaves are located within a dedicated portion of a processor's memory --
this is known as the {\em Enclave Page Cache (EPC)}. 
The EPC is a continuous block of memory (128-256 MB) that is initialized during the boot process.
It is inaccessible to the operating system and hypervisor. 
Some of its contents (excluding the metadata) are accessible to processes running within it 
(with appropriate isolation between secure processes themselves).
In SGX, only the on-chip components such as the
processors, caches, NoC and memory encryption engine (MEE) are assumed to be secure. These secure components are a part
of the  {\em trusted computing base (TCB)}.

\begin{figure}[!htb]
    \centering
    \includegraphics[width=0.6\linewidth]{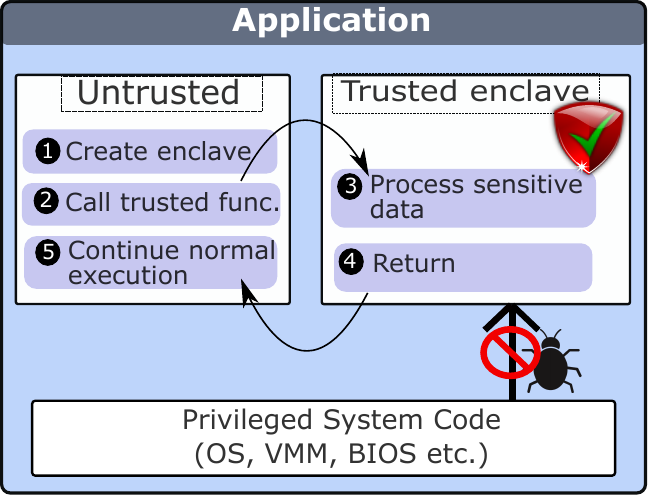}
    \caption{SGX runtime execution}
    \label{fig:sgx}
\end{figure}

Enclaves are created before invoking the trusted code during execution. When we call a
trusted function, secure execution starts
within the enclave. Once the execution completes,
the function returns and the context is switched back.
Subsequently, normal unprotected execution of the application continues
(refer to 
Figure~\ref{fig:sgx}).
 
The OS manages the page tables and the TLBs. However, any update to the TLB needs to be 
vetted by the SGX 
subsystem. Hence, a dedicated HW circuit
verifies the integrity of the contents of the secure page and also ensures that no ``secure'' virtual
address is mapped to an ``unsecure'' physical page or vice versa using an inverted page table. 
When the EPC is full, we need to evict a page, encrypt it and store it in the unsecure part of memory. 
Some metadata corresponding to it such as the key used to
encrypt it and its MAC (keyed hash) are stored in the EPC. To reduce storage space, we can create
an eviction tree of such evicted pages that is similar to the classical Merkle tree.
Note that entering and exiting secure mode are expensive operations ($\approx$
20-40k clock cycles \cite{vault,sgx}). So is bringing back an evicted page to an EPC, hence, 
{\bf EPC misses should be minimized.}

\subsection{SGX-Client: (\texorpdfstring{$10^{th}$} Gen. Intel CPUs)}
\label{sec:client}
SGX-Client employs a memory encryption engine (MEE) \cite{mee}, which is 
an extension of the memory controller (part of TCB).
To maintain the {\em confidentiality} of the data, the
MEE encrypts the data using {\em Advanced Encryption Standard (AES) counter-mode encryption\cite{aesctr,ctr} (AES-CTR)}. The counter values correspond to different data blocks of a page; 
whenever a data block is modified, its counter is incremented by one (to stop replay attacks). 
The inclusion of these counters in the encryption processs guarantees that \underline{effectively a new key} 
is used for every encryption of the same block.
The {\em integrity} of the counters is essential to the system as their correctness directly affects the system security. Their {\em integrity} is ensured in SGX via MACs stored in memory
and a {\em Merkle tree} that aggregates them. 

\subsubsection{Integrity Verification using Merkle Trees} 
The leaf nodes of the Merkle tree store counters for the 
secure pages (part of the EPC), and the internal nodes of the tree store the
counters for each of their child nodes. Additionally, each node 
stores the MAC of its counters (encrypted hash), in such a way that a
{\em Carter-Wegman}\cite{carter-wegman} style tree is created --
the MAC is generated by encrypting the hash of the counters in the node
using the counters in its parent node. 
We can thus conclude that the root node
captures the information of all the nodes in the tree. If there is a change in any counter,
it will get reflected at the root. We thus need to store the root of the tree in the TCB
and for efficiency we can store additional nodes of the tree in the  TCB such that the
root need not be updated on every write.

As we increase the size of secure memory, the number of counters increases, thereby increasing 
the size (depth) of the Merkle
tree and its associated storage overheads.
As the depth of the tree increases, the integrity verification of the
counters starts taking a lot of time. 
Hence, the size of the EPC has been limited to 128MB or 256MB in
SGX-Client. 
Sadly, a part of the EPC memory has to be
reserved for storing such EPC metadata, which further decreases its size (32 MB out of 128 MB). 
\vspace{-2mm}
\subsection{SGX-Server: (\texorpdfstring{$11^{th}$} and \texorpdfstring{$12^{th}$} Gen. Intel CPUs)}
Intel launched the $3^{rd}$ Gen. Intel Xeon Processor based server platforms in 2021
to facilitate trusted execution environments (TEEs) that support a
large number of enclaves and a large secure memory (SGX-Server). SGX-Server
uses TME-MK (Total Memory Encryption-Multi Key) \cite{tmemk}
that uses multiple keys to encrypt the physical memory -- one key for each VM. 

In SGX-Server,
the physical memory is encrypted using the AES-XTS (Advanced Encryption Standard -- Tweakable Block
Ciphertext Stealing) \cite{xts} encryption engine. 
AES-XTS is used for block-based storage devices and takes into
account the physical address for encrypting a data block. 
For encryption, AES-XTS uses a 128-bit tweak derived form
the physical address to provide address-based variability and the Galois Function is applied on the encryption engine output to add an element of
diffusion. 

The AES-XTS encryption lies on the critical path and encrypts all the data that
enters or leaves the chip. Given that we have confidentiality and integrity (via MACs),
several hardware attacks (like cold boot attacks) and memory bus probing or relocation/splicing
attacks are prevented. Even though, it is much faster and more scalable, a key security guarantee
is sacrificed -- replay protection or freshness\cite{freshness}. This means that it is possible to replace
the value in a memory location (along with its MAC) with a pair of values that were seen in the past.
The processor will not be able to perceive that the memory contents have been tampered with. Also,
it is possible to say that the value in a memory location is the same as that at a previous point
of time -- the ciphertext will be the same (this is a side channel).
\vspace{-3mm}
\section{Threat Model}
\label{sec:threat}
\begin{figure}[ht]
    \centering
    \includegraphics[width=0.6\linewidth]{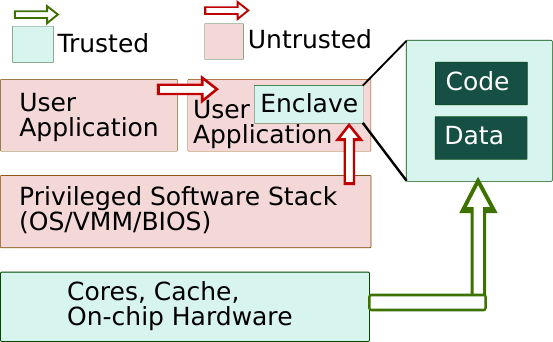}
    \caption{Threat model}
    \label{fig:tm}
\end{figure}

\fname considers a threat model similar to that of Intel 
SGX~\cite{sgx}.
We include only the on-chip hardware components in the TCB, which includes the
cores, caches, NoC, MEE and the hardware circuits that we introduce in \fname.
Other than these components, we only trust the code running within the enclaves with regards
to their own execution. 
The SGX enclaves and standard cryptographic operations maintain confidentiality and detect integrity violations. The hardware components outside the TCB, the privileged software stack and other user applications including unrelated enclaves, are considered to be untrusted~\cite{graphene}. Figure~\ref{fig:tm} displays the trusted and untrusted components in the system.

Similar to \cite{vault,morph}, \fname assumes that the attacker controls 
the {\em system software stack} and can misuse
its privileges to launch attacks such as observing and modifying the contents
of memory addresses~\cite{system_software}. 
The attacker can also mount physical attacks such as snooping on the
memory bus or cold boot attacks -- observe and modify any memory location at will\cite{cba}. 

\fname provides all four ACIF guarantees: authenticity, confidentiality,
integrity and freshness.
It protects the system against {\em replay attacks}\cite{replay}, where the adversary may 
replace a data-block/MAC pair in memory.
Akin to SGX and similar TEE schemes, \fname does not consider side
channel attacks (power, EM and cache), DoS attacks and attacks that introduce errors
in the computation based on laser pulses or voltage spikes~\cite{plunder}.

\vspace{-5mm}
\section{Characterization}
\label{sec:char}

The aim of characterization is to determine the sources of performance degradation in SGX systems by characterizing the behavior of
benchmarks and baseline designs. 

\subsection{Setup and Benchmarks} 
We ran the SPEC CPU 2017\cite{spec} benchmarks and characterized 
their performance on different systems
(standard practice while evaluating TEEs). The different systems are modeled and simulated in a cycle-approximate simulator, Tejas\cite{tejas}. 
Table \ref{tab:specs} shows 
the simulation parameters. 
We used an algorithm similar to PinPoints~\cite{pin_tool} and SimPoints~\cite{simpoint} to find the regions
to simulate in each workload, and then we weighted them appropriately~\cite{weighted} to arrive 
at the final figures. We used Intel Pintools 3.21 \cite{pin}.

\begin{table}[!htb]
\footnotesize
    \centering
    \begin{tabular}{ |c|c|c|c| }
    \hline
    \rowcolor{blue!10}
    \multicolumn{4}
       {|c|}{\textbf{Processor}} \\
    \hline
    \hline
    \textbf{Parameter} &\textbf{Value} &\textbf{Parameter} &\textbf{Value} \\
    \hline
    Cores& 1 & Pipeline & 4 Issue \\
     & & &Out Of Order \\ \hline
    \hline
    \rowcolor{blue!10}
    \multicolumn{4}{|c|}{\textbf{Caches}} \\ \hline
    \textbf{Cache}&\textbf{Size}&\textbf{Type}&\textbf{Associativity} \\
    \hline
    L1 I-cache& 32 KB & Private & 8 \\ \hline
    L1 D-cache& 32 KB & Private & 8 \\ \hline
    L2 cache& 8 MB & Shared & 8 \\ \hline
    Counter cache& 32 KB & Shared & 1 \\ \hline
    \hline
    \rowcolor{blue!10}
    \multicolumn{4}
    {|c|}{\textbf{Memory}} \\ \hline 
    \hline
    \textbf{Parameter}&\textbf{Value}&\textbf{Parameter}&\textbf{Value} \\
    \hline
    Frequency& 3.6 GHz & Channels & 2 \\ \hline
    Ranks & 2 & Banks & 8 \\ \hline
    Ports & 1 & Port Type & FCFS \\ \hline
    \hline
    \end{tabular}
    \caption{System Specifications}
    \label{tab:specs}
\end{table}
\vspace{-3mm}
\subsection{Systems Modeled}
\circlenew{1} \textbf{Baseline} is a vanilla design
with no security. 
\circlenew{2} \textbf{SGX} is SGX-Client that implements a Merkle Tree and a
128MB EPC. It guarantees all ACIF properties.
\circlenew{3} \textbf{DFP} has a Merkle tree and a 128MB EPC, which
is the same as baseline SGX. 
It additionally implements a predictor that 
predicts page faults in the near future
and prefetches the pages. \circlenew{4} \textbf{Penglai} has a Mountable Merkle Tree 
(with a root tree and multiple
subtrees) that can support 512GB of secure memory, but does not have an 
EPC. It only caches the 32 most recently seen subtree roots.
For every LLC miss, multiple additional memory accesses are required to retrieve
the tree nodes for integrity verification.

\begin{scriptsize}
\footnotesize
\begin{table}[!htb]
    \centering
    \begin{tabular}{|l|l|l|}
    \hline
    \rowcolor{blue!10}
         \textbf{System}&\textbf{Integrity Tree}  &\textbf{EPC} \\
         \hline
         \hline
         Baseline& No & No \\
         \hline
         SGX-Client & Merkle Tree & 128MB EPC \\
         \hline
         DFP& Merkle Tree & 128MB EPC \\
         \hline
         Penglai& Mountable Merkle Tree & No \\
         \hline
    \hline
    \end{tabular}
    \caption{Security Constructs of the Models Simulated}
    \label{table:models}
\end{table}
\end{scriptsize}
\vspace{-4mm}
\subsection{Observations}

\subsubsection{Performance Comparison}
\begin{figure}[!htb]
    \centering
    \includegraphics[width=0.9\linewidth]{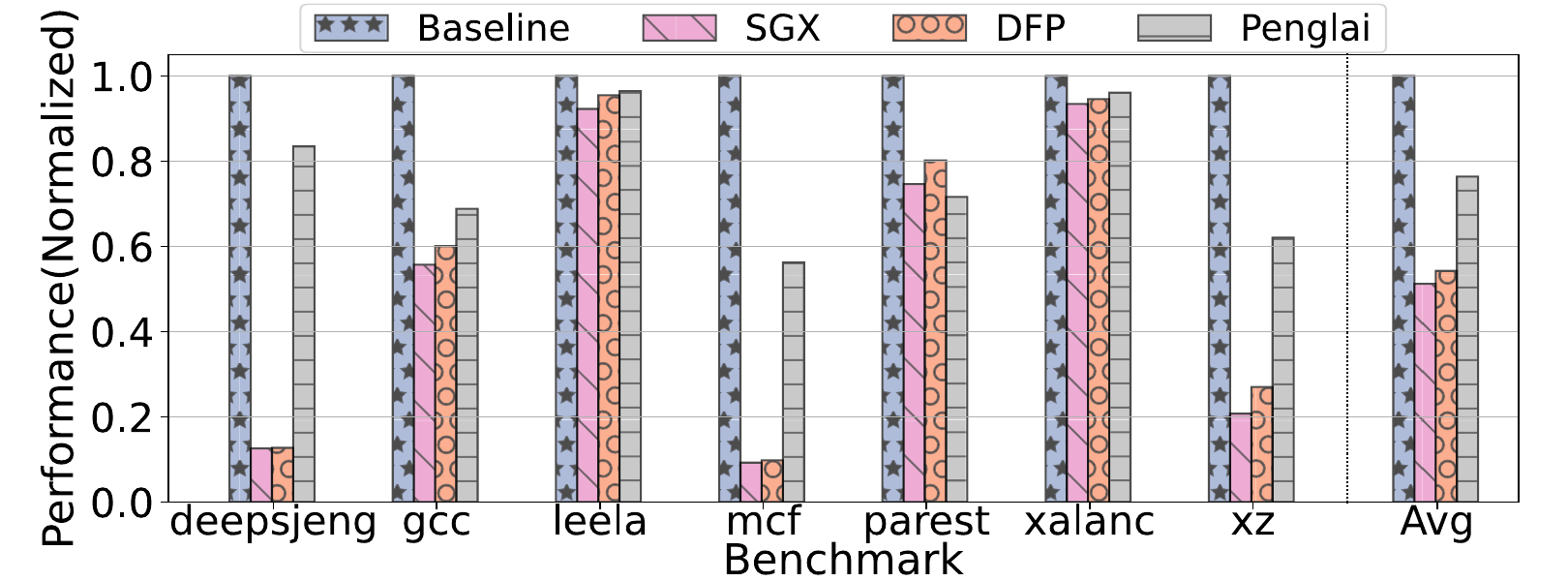}
    \caption{Comparing the performance of different secure architectures}
    \label{fig:ipc_char}
\end{figure}

The performance (reciprocal of the simulated execution time) of
the systems for
different workloads is shown 
in Figure \ref{fig:ipc_char}. In comparison to baseline, SGX-Client shows
a very high performance degradation (mean: 83\%) due
to the overheads associated with traversing the Merkle tree and the EPC page fault penalties.
DFP shows very little improvement in performance ($\approx$ 2\%), compared to SGX-Client. 
It is limited by the accuracy of its EPC page fault predictor. 
Penglai shows better performance than SGX-Client and DFP (mean: 49\% better than SGX-Client).
The source of its overheads is the latency incurred due to additional memory accesses to the MMT (Mountable Merkle Tree) required for integrity verification. 

\subsubsection{Sources of Performance Overheads}
\begin{figure}[!htb]
    \centering
    \includegraphics[width=0.8\linewidth]{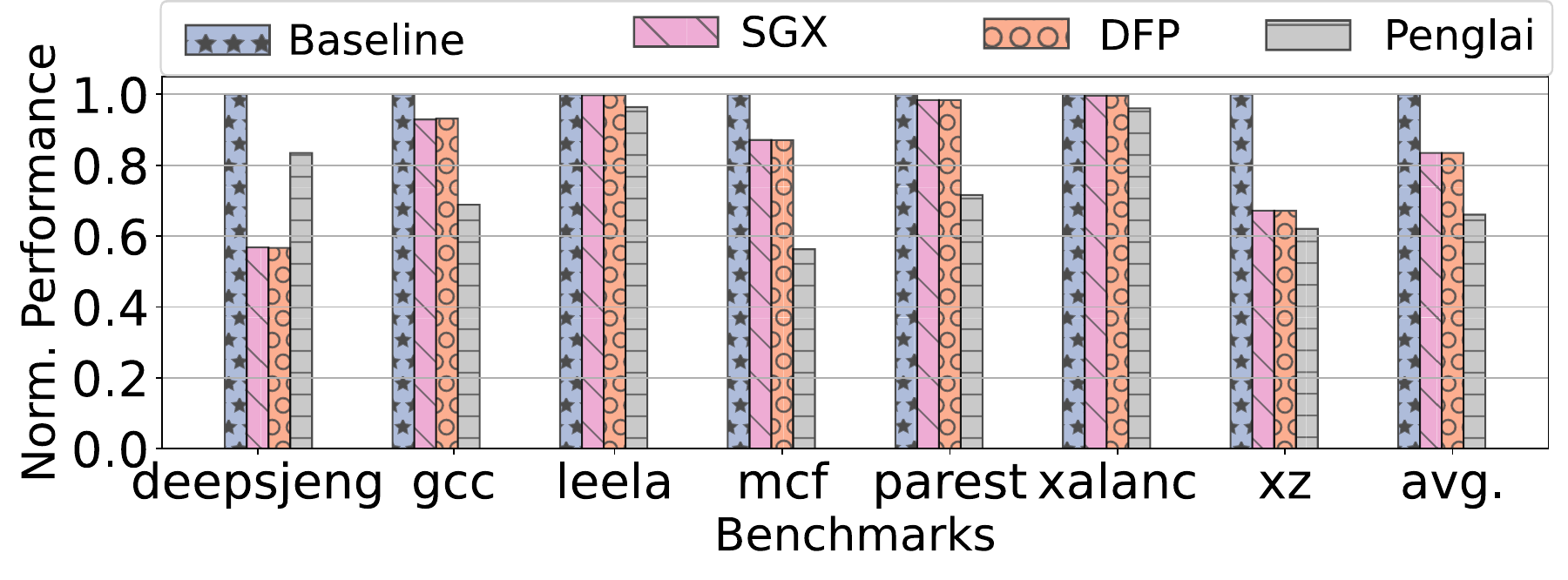}
    \caption{Performance with only the Merkle tree/MMT traversal/maintenance overhead. Results: 17$\%$ performance degradation is observed in SGX and DFP, whereas a 34$\%$ dip is observed in Penglai w.r.t. the baseline model.}
    \label{fig:mer}
\end{figure}

Let us separately analyze the impact of Merkle tree traversal and EPC page fault 
servicing.  Figure~\ref{fig:mer} shows the performance of the different systems
with only the integrity tree (Merkle Tree/MMT) overheads. We assume a
zero EPC page fault penalty.  

There is a degradation in performance observed for all the systems. However, the performance degradation in Penglai is
observed to be the highest (34$\%$). This is because its integrity tree, MMT, is larger
than that of the other two
systems. It supports counters for 512GB memory. Additional memory accesses are required for every secure memory access
to retrieve the counters and verify their integrity.
\textbf{Conclusion: Encrypting large secure memory using counter-mode encryption with an integrity tree (Merkle Tree/MMT) fails to scale
well, it rather imposes large performance overheads.}

\begin{figure}[!htb]
    \centering
    \includegraphics[width=0.8\linewidth]{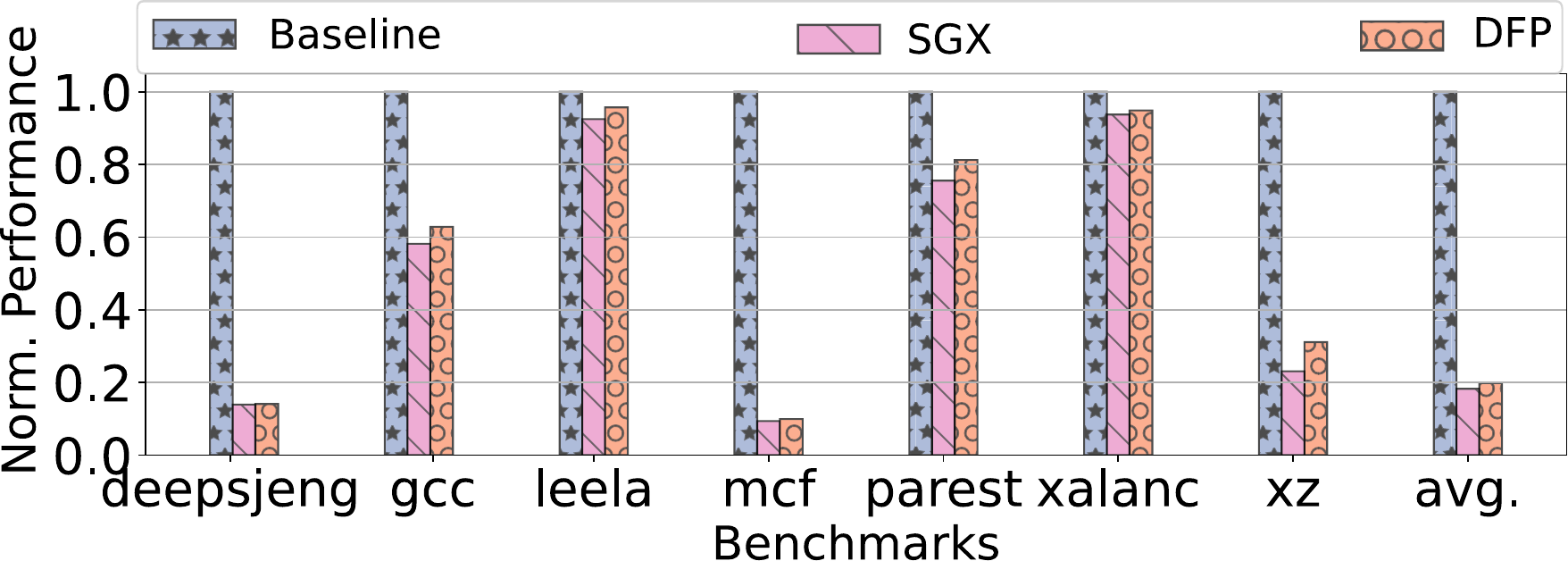}
    \caption{Performance with only the overhead of EPC page faults.
    Results: Performance degradation w.r.t. baseline is observed to be 82$\%$ in SGX and 80$\%$ in DFP.}
    \label{fig:epc}
\end{figure}

Figure \ref{fig:epc} shows the effect of overheads due to EPC page fault handling.
We assume the overheads associated with the integrity trees to be zero.
We observe that the performance degradation is much more drastic because 
of these overheads. The existing page fault handling mechanism is 
very costly and takes a large number of cycles ($\approx$ 40k cycles \cite{vault}) to complete.
The entire page loading process is slow because of all the DRAM reads, decryption and
updation of metadata -- this
increases the length of the critical path.

SGX-Client and DFP suffer from overheads resulting from
both the Merkle tree as well as EPC page faults. 
DFP shows little improvement over SGX-Client ($\approx$2\%), 
owing to preloading of pages in case of correct predictions.
However, the prediction accuracy of DFP varies significantly across benchmarks
(0-19.7\% in our experiments). 
As Penglai does not have a limited-size EPC, it does not suffer from 
overheads associated with EPC page faults; however, maintaining  the integrity information
is quite onerous in its case.
%FIXME: Isn't the prediction accuracy of DFP quite low?

\subsubsection{Analyzing EPC Page Fault Overheads}
\begin{figure}[!htb]
    \centering
    \includegraphics[width=0.9\linewidth]{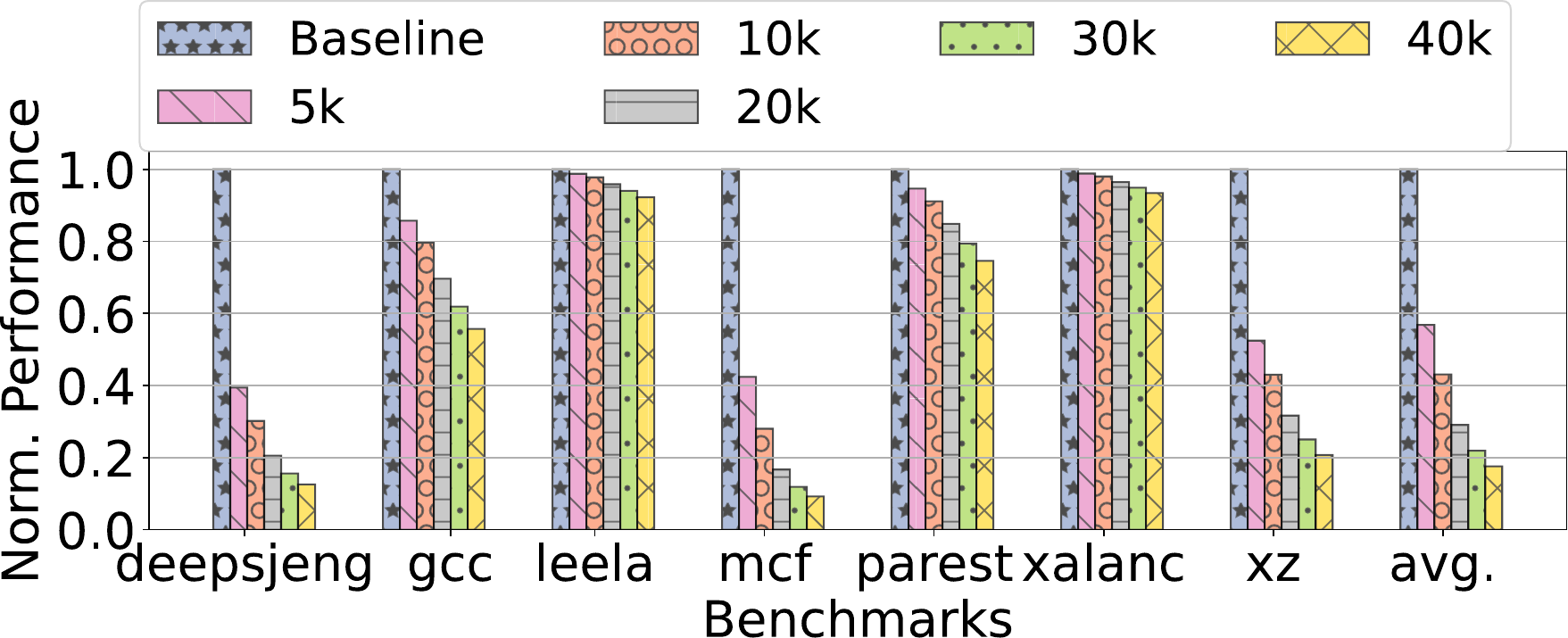}
    \caption{Comparing the performance for varying page fault penalties in SGX.
    Baseline is the unsecure system, 5k, 10k, 20k, 30k and 40k refer to the variation in EPC page fault penalties, respectively (unit: clock cycles). 
    \textit{Conclusion: The page fault penalty directly increases the latency of the \textbf{critical path}.} }
    \label{fig:ipc_var}
\end{figure}

To analyze the impact of EPC page fault penalties on the performance of the system, we simulate the system with varying values of the EPC page fault penalty 
and observe the difference in the performance (see Figure \ref{fig:ipc_var}).
The EPC page fault penalty
has a direct impact on system performance as it directly affects the
latency of the critical path. The overhead increases from 44\% (5k cycles) to 83\% (40k cycles)
relative to the baseline.

\begin{figure}[!htb]
    \centering
    \includegraphics[width=0.7\linewidth]{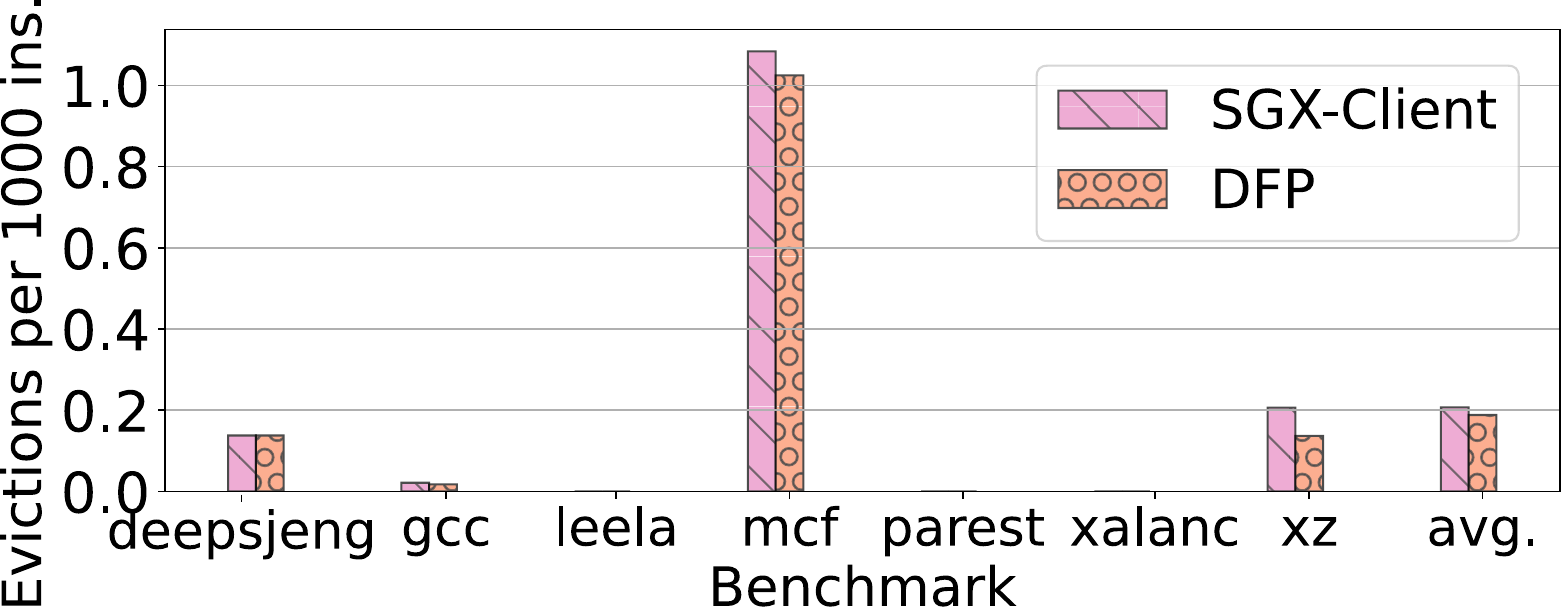}
    \caption{Evictions per 1k instructions for different workloads}
    \label{fig:evictI}
\end{figure}

\vspace{-2mm}
Since EPC page faults result in large performance overheads,
we analyzed the frequency of such events by computing the number of evictions per
1000 instructions in different workloads (see Figure \ref{fig:evictI}).
We observe that the value is quite low in most cases:
on an average {\em 0.2 evictions per 1k instructions}. 
\textbf{Although infrequent, these page faults have a huge impact on the system performance due to their excessively high latency.} 

\subsubsection{Storage Overheads}

\begin{figure}[!htb]
    \centering
    \includegraphics[width=0.7\linewidth]{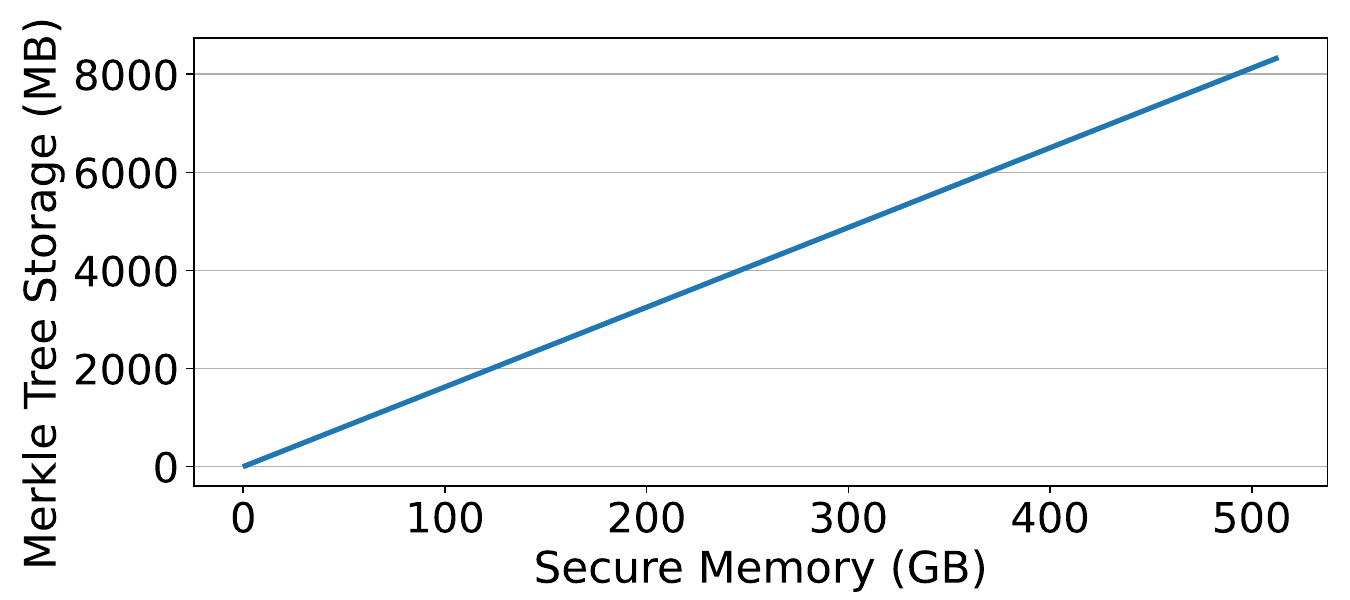}
    \caption{Storage overhead of using counters for guaranteeing freshness}
    \label{fig:counter_overhead}
\end{figure}

In addition
to the performance overheads that we saw earlier, 
the additional storage overhead can be visualized in Figure
\ref{fig:counter_overhead}. The overhead varies linearly.
Over 8 GB of memory is required to store the counters for
512 GB memory. Extending the Merkle Tree to add a leaf node counter (for a page in secure memory) may require the addition of multiple nodes in the tree (parent nodes). 
\textbf{Unrestricted scalability with freshness guarantees can only be achieved if key management adds modest storage overheads.} 
There is a need to devise a more efficient mechanism for providing freshness guarantees
that scale to TBs of physical memory.

\begin{table}[hbt!] %[!htb]
    \centering
    \begin{tabular}{|p{0.95\columnwidth}|}
     \hline
       \rowcolor{gray!20}
         \textbf{Insight 1} Encrypting large memory using counter-mode encryption with a Merkle Tree does not scale well and imposes large performance overheads as the size of secure memory increases. \\
       \rowcolor{gray!20}  
         \textbf{Insight 2} The huge latency of EPC page fault management is the reason why merely 0.02$\%$ of instructions (that cause EPC page faults) lead to about 82$\%$ performance reduction in SGX. Reducing the wait time associated with this latency is the key to reducing the latency of the critical path. \\
       \rowcolor{gray!20}  
         \textbf{Insight 3} The management of keys 
         to provide freshness guarantees should not lead to large storage overheads.
         \\
      \hline
    \end{tabular}
\end{table}
\vspace{-3mm}
\section{System Design}
\label{sec:design}

\subsection{Overview}

\begin{scriptsize}
\footnotesize
\begin{table}[!htb]
    \centering
    \begin{tabularx}{\columnwidth}{|X|}
    \hline
    \rowcolor{blue!10}
         \textbf{Insight 1: Merkle tree restricts scalability.}  \\
         \hline
            Employ counter mode encryption along with a Merkle tree only for pages in the EPC (like SGX-Client). Implement a
            different kind of encryption for the eEPC region to ensure smooth scalability of enclaves. \\
         \hline
         \hline
    \rowcolor{blue!10}
         \textbf{Insight 2: EPC page fault management latency needs to be reduced.}  \\
         \hline
            Implement the {\em read first, verify later} approach using a separate MAC verification circuit and perform security operations concurrent to speculative execution. Reduce the granularity of the page load/eviction process to reduce the wait time.\\
         \hline
         \hline
    \rowcolor{blue!10}
         \textbf{Insight 3: Freshness guarantee for scalable system demands effortless key management strategy. }  \\
         \hline
            Use randomly generated page-level keys for encryption and secure the keys in protected memory using encryption and MACs. An efficiently designed MAC forest protects the integrity of this memory. \\
    \hline
    \end{tabularx}
    \caption{Design decisions based on insights derived from characterization (Section \ref{sec:char}).}
    \label{table:char_design}
\end{table}
\end{scriptsize}

\begin{figure}[!htb]
    \centering
    \includegraphics[width=0.7\linewidth]{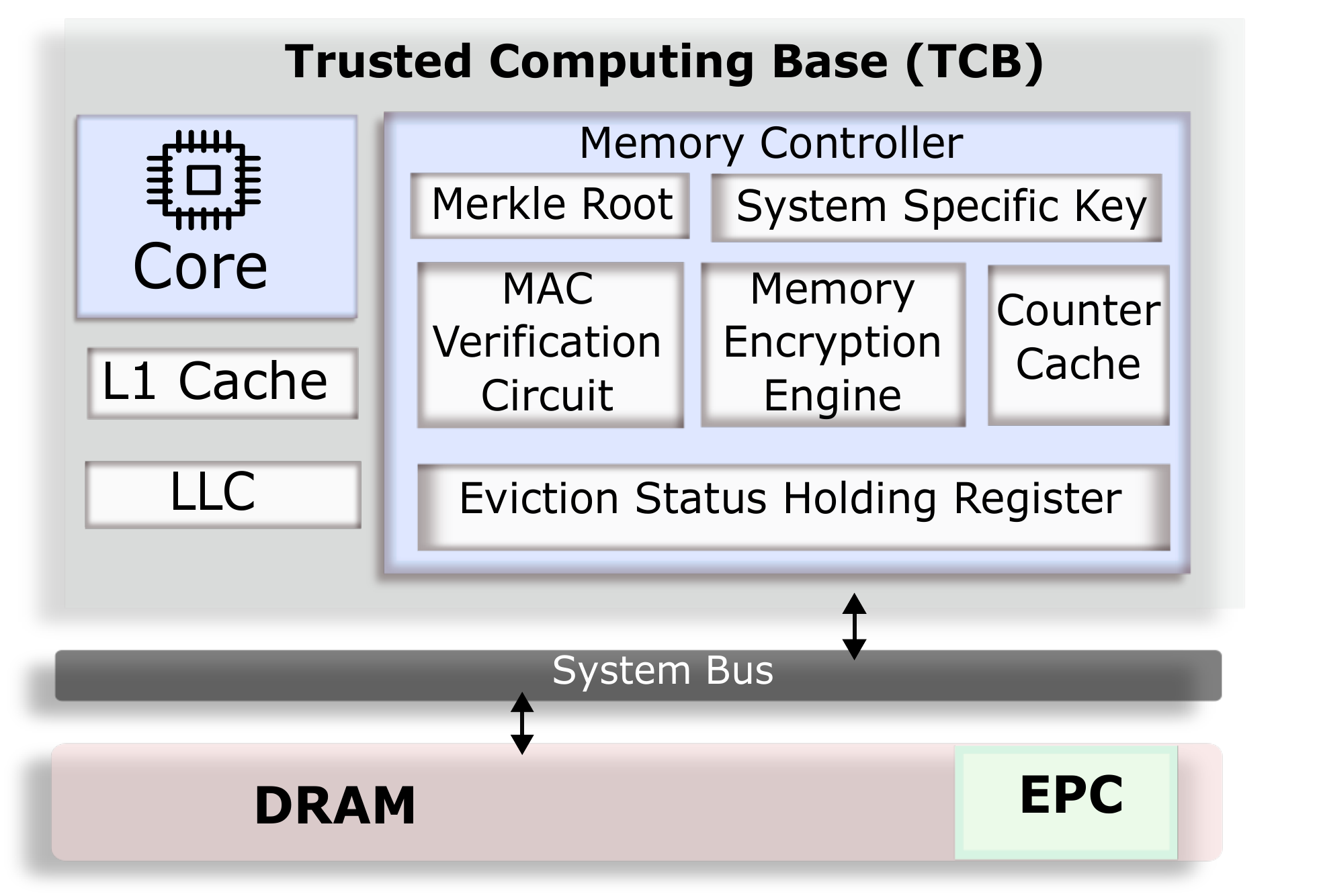}
    \caption{The high-level design of \fname }
    \label{fig:arch}
\end{figure}

The design principle of \fname revolves around the {\em read first, verify later} paradigm.
Our system supports enclaves that are
capable of handling large workloads up till 512 GB (similar to SGX-Server);
the users get full ACIF security (similar to SGX-Client).
In the basic design, an unlimited number of enclaves are supported (total size:  512 GB).
This is achieved through efficient utilization of the pre-existing 128 MB EPC (part of 
SGX-Client)
to create an eEPC (extended EPC) region of 512 GB (hereby, named eEPC). 
It provides a hardware-assisted secure execution environment for server applications. 
The design decisions are summarized in Table \ref{table:char_design} (based on the
characterization).
The high-level design of \fname is shown in Figure~\ref{fig:arch}.  

\subsection{MAC Forest for Integrity Verification}

Taking a more efficient approach towards integrity protection, we propose a hierarchical MAC-forest-based integrity mechanism to secure the entire eEPC region.
In the eEPC region of \fname, MACs are computed at the page-level rather than block-level. A MAC engine (ME)
computes an 8-byte MAC for each 4 KB page. These MACs are stored as the leaf nodes in the MAC forest. Based on the arity
of the individual subtrees in the MAC forest, we group {\em p} MACs from the leaf nodes to generate a single 8-byte parent MAC. This process is
repeated at all the levels of the subtrees. 
A part of the eEPC is reserved to store the lower level nodes of the subtrees in the MAC forest.
and the topmost level of these subtrees is securely stored in the EPC (part of the TCB). 

We retain the use of counters and the Merkle tree for protecting the EPC region (akin to SGX-Client) with the same permission scheme as SGX-Client for accesses.
In our evaluated design, we consider a MAC forest consisting of subtrees with {\em q} $=$ 3 (3-level tree) and {\em p} = 16 $\times$ 8 (arity 16
at the lower level and 8 at the higher level). In this case, the MACs of 16 pages will be grouped in the lower level to
generate a MAC at the parent. For the next level, we group 8 MACs to generate a parent MAC which is part of the top most level of the forest. For 512 GB memory, 
we get a MAC forest
with the topmost level containing $2^{20}$ MACs (securely stored in the EPC, 8 MB total).
By limiting the levels of the subtrees we can contain the
additional memory bandwidth required for integrity verification (maximum of 4 additional accesses in our representative
design).

\begin{figure}[!htb]
    \centering
    \includegraphics[width=1.0\linewidth]{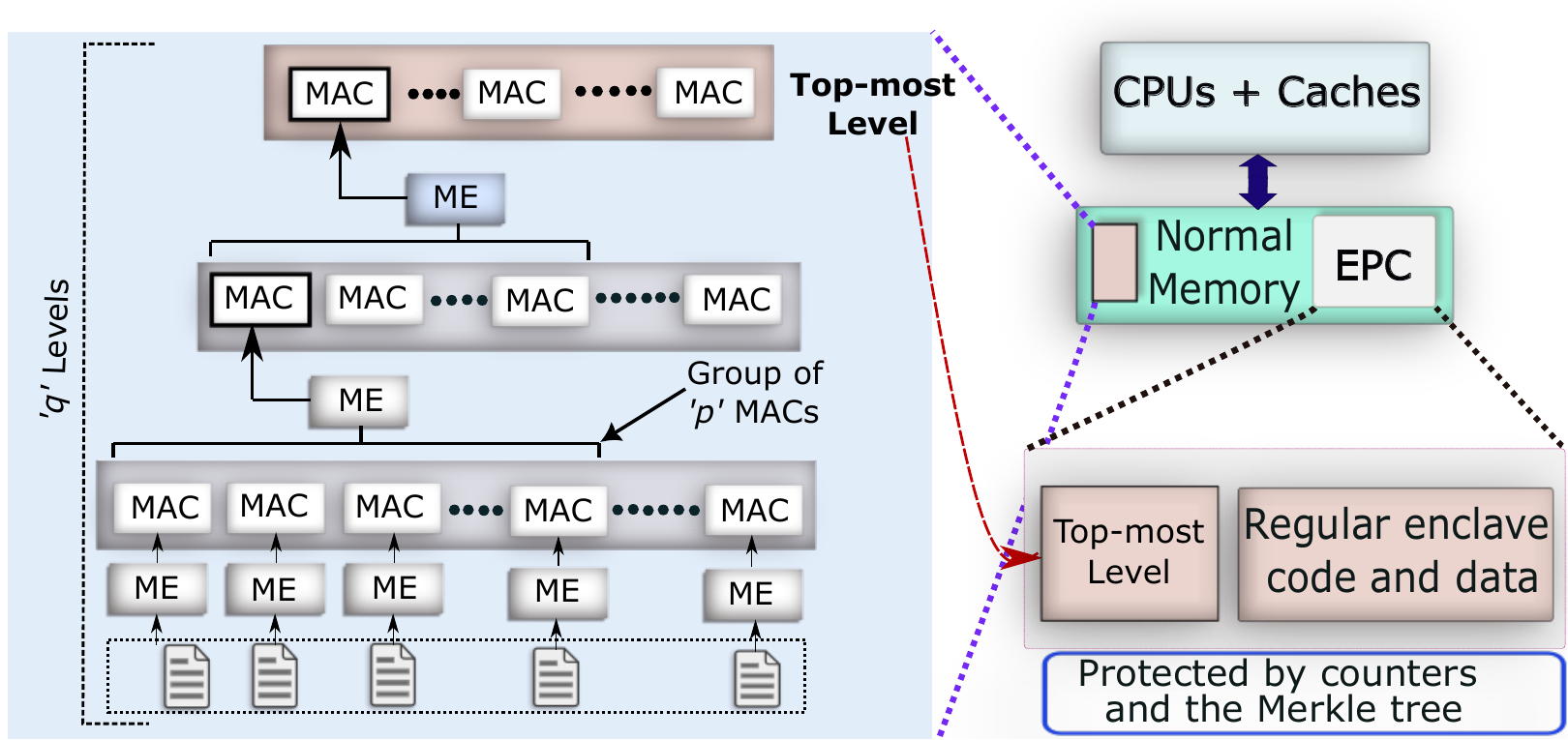}
    \caption{MAC forest with numerous subtrees. (ME refers to the MAC engine)}
    \label{fig:scalable_scheme}
    \vspace{-4mm}
\end{figure}

\noindent \textbf{MAC Verification Circuit (MVC):} We implement a MAC verification circuit (MVC) that is responsible for verifying the integrity of the pages within the eEPC region. 
The MVC performs this verification when a page is loaded into the EPC.
Note that we perform \textit{deferred MAC verification} -- we do not halt the normal execution to wait for the outcome of the verification process. It uses the SHA-2 algorithm to compute MACs (Throughput: 40 Gbps at 5.15 GHz frequency and 7 nm tech. node).

\begin{table}[hbt!] %[!htb]
    \centering
    \begin{tabular}{|p{0.95\linewidth}|}
     \hline
       \rowcolor{gray!20}
         \textbf {\em The Merkle Tree and counters are limited to the EPC (like SGX-Client) and the MAC Forest is used for the eEPC region (with MVC).} \\
      \hline
    \end{tabular}
\end{table}
\vspace{-6mm}
\subsection{Full Memory Encryption}
\label{sec:fme}
To protect the pages in the eEPC region we use vanilla AES-ECB encryption with a few tweaks. 
Every page in the eEPC region is encrypted using a different encryption key
that is randomly generated every time a modification is made.
This mechanism effectively safeguards against replay attacks because keys are not reused (probabilistically). 

For AES-ECB mode encryption, we use a 256-bit key, which is a combination of a hardware-specific (HW) key (64 bits), enclave ID (31 bits),
bits generated by a pseudo random number generator (128 bits) seeded by the boot time and HW key,
physical address of the page in 512 GB memory (27 bits) 
and block address within the page (6 bits). We keep the PRNG component large
because a new value needs to be created for every encryption.
The components of the key excluding the block-specific 6 bits constitute the page specific key, \textit{K}. 
The block-specific address bits are extracted for every block of the page and concatenated with \textit{K} to generate the block-specific key $k_b$, where $b$ represents the $b^{th}$ page block.
This ensures that the keys used for encrypting each block of the page are different. Thus, 
if the same data is stored in different memory blocks of the page, different ciphertexts will be generated. 

\begin{figure}[htb!]
    \centering
    \includegraphics[width=0.8\linewidth]{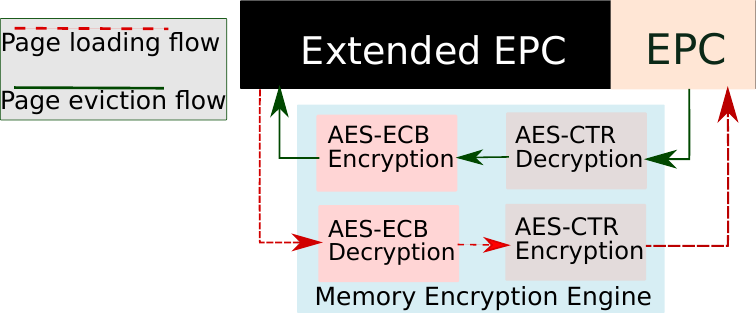}
    \caption{Encryption-decryption during EPC page eviction and loading}
    \label{fig:ec1}
\vspace{-3mm}
\end{figure}

\subsubsection{EPC -- eEPC Page Transfer} 
The encryption-decryption process when a page is getting transferred from the EPC to the
eEPC is shown in Figure~\ref{fig:ec1}.
It is triggered by an EPC page fault.
When a page is evicted from the EPC, the encrypted page is first decrypted using the AES-CTR mode, and then it is re-encrypted using the AES-ECB-256 mode with a secret key generated by the MEE. 

\begin{table}[hbt!] %[!htb]
    \centering
    \begin{tabular}{|p{0.9\linewidth}|}
     \hline
       \rowcolor{gray!20}
         \textbf {\em In the eEPC region, \fname generates a new key every time the page is written to. This guarantees freshness. }\\
      \hline
    \end{tabular}
\end{table}
%FIXME: Replace extended EPC with eEPC

The page specific key, $K$, is then encrypted using a system-specific key {\em SSK}. The {\em SSK} is
a concatenation of a
second device-specific key (128 bits) and the 128-bit boot time. The {\em SSK} is stored in a dedicated register in the TCB. 
The block-specific values in the key are extracted from the address at the time of cryptographic operations, and only
the randomly generated page specific key, \textit{K}, is stored in the eEPC region.  When encrypting and sending the
key to the eEPC, the block address within the page is set to zero.  The encrypted page
and its encrypted key are then moved to the eviction region -- both are stored in the eEPC region.

A section of the physical memory called the Key Table
stores all the encrypted keys for each of the physical pages. Given that
we have $2^{39-12}$ (=128M) physical pages in our system and each key is 16 bytes, we need 2GB of storage
for storing the keys. This translates to 0.4\% overhead for storing the keys in physical memory.
The advantage of storing the keys in this manner is that we can easily
locate the key given the page's physical address and fetch it along with the evicted page 
when it is required in the EPC. Basically, an evicted page comes to the EPC along with its encrypted key.
We trust {\em the key  for the time being} till verification. Thus, key storage and management
overheads are effectively reduced while maintaining key freshness.  While bringing a page from the
eEPC to the EPC region, a reverse process is followed (as shown in Figure \ref{fig:ec1}). First the page is
decrypted using the AES-ECB mode using the encrypted key read from the Key Table
and then its blocks are re-encrypted using the AES-CTR mode (within the EPC).

\begin{table}[hbt!] %[!htb]
    \centering
    \begin{tabular}{|p{0.95\linewidth}|}
     \hline
       \rowcolor{gray!20}
         \textbf{\em The integrity of the keys in the Key Table are maintained as follows: 
         Use the page-key to encrypt the hash of the page and generate the MAC. The key cannot be unilarerally changed (the MAC check will fail), the key and the MAC cannot be replayed (the MAC check at the higher level will fail) and another enclave cannot read or write the data (the MAC check will fail). Higher-level MACs are created by encrypting the hash with the SSK.} \\
      \hline
    \end{tabular}
\end{table}
\vspace{-3.5mm}
\subsection{Optimization the EPC Page Fault
Handling Mechanism}
\begin{figure}[!htb]
    \centering
    \includegraphics[width=0.8\linewidth]{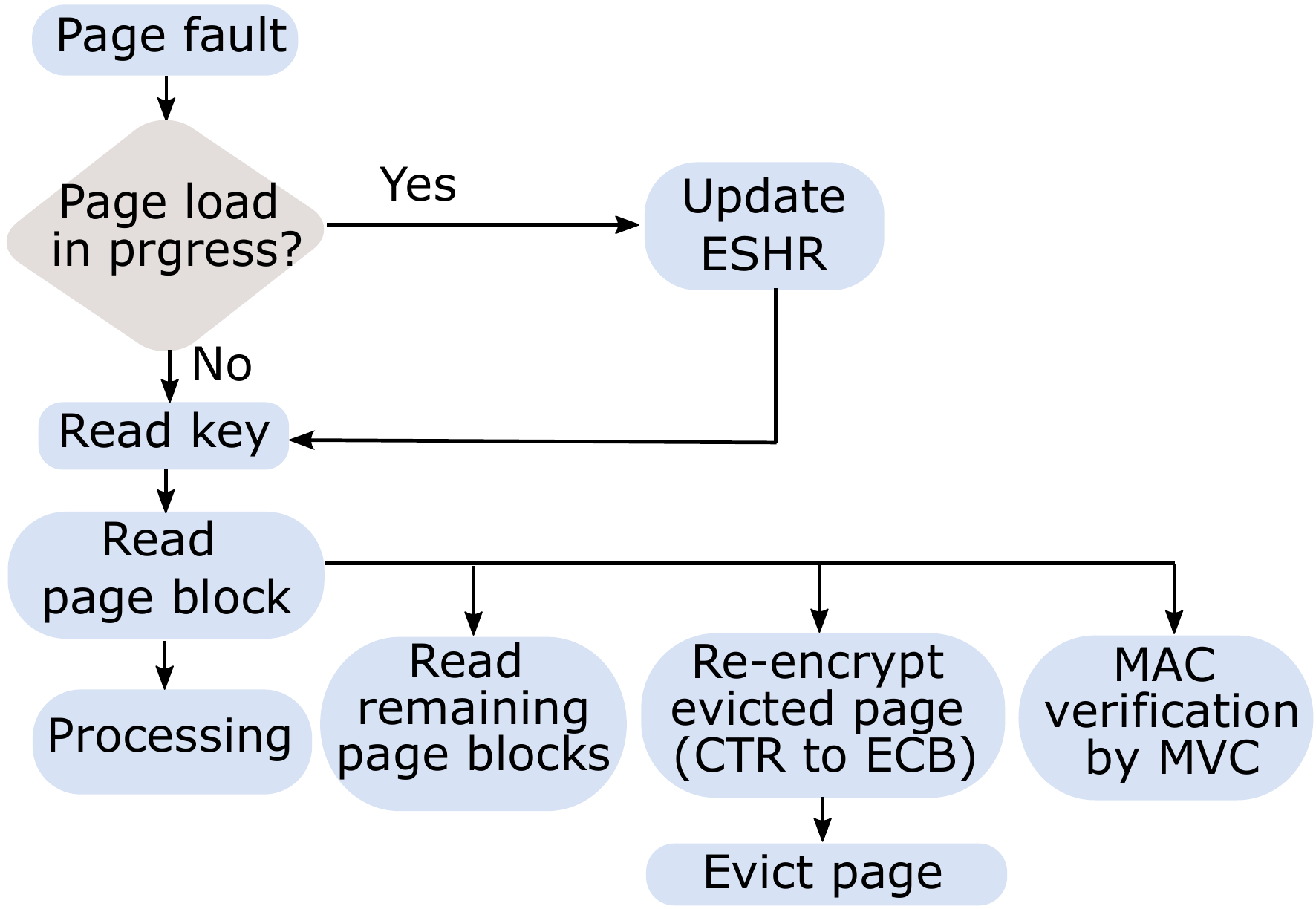}
    \caption{Handling EPC page faults}
    \label{fig:flowchart}
    \vspace{-4mm}
\end{figure}

Figure~\ref{fig:flowchart} shows the entire process of fetching a page
along with its key, decrypting and verifying it. 

If there is no other concurrent EPC miss, 
this process continues without interruption. Assume another EPC page fault
occurs before this page is fully loaded
(i.e., before all the blocks of the mapped page have been loaded into the
EPC). The current process of loading the EPC page
must be run {\em in parallel} with processing the new request.
We need to first store the status of the ongoing
eviction/loading process so that loading can be resumed later from the current state. 
We introduce an
additional hardware structure called the ESHR Table to store this
information regarding the page loading and eviction status. Each entry in the table
is an Eviction Status Holding Register (ESHR).
Subsequently, the block of the new page along with its
key are fetched into the EPC. Once it is fully loaded we resume the process of loading
the rest of the blocks of the page whose status was saved in an ESHR. 

\noindent \textbf{ESHR Table} 
To keep track of which page blocks are loaded in the EPC, 
we maintain a 64-bit load status vector ($LS$ vector) in the
ESHR. We have 64 bits because there are 64 blocks in a page.
When a block is loaded into the EPC, its corresponding bit in the $LS$ vector is set to 1. 
Once all the blocks are
loaded, the valid bit ($V$) is reset.  
The required page ($LPage$) is loaded in place of a corresponding evicted page
($EPage$) as indicated in the ESHR. The eviction bit ($E$) is set to 1 if there is eviction along with loading.

\begin{figure}[!htb]
    \centering
    \includegraphics[width=0.9\linewidth]{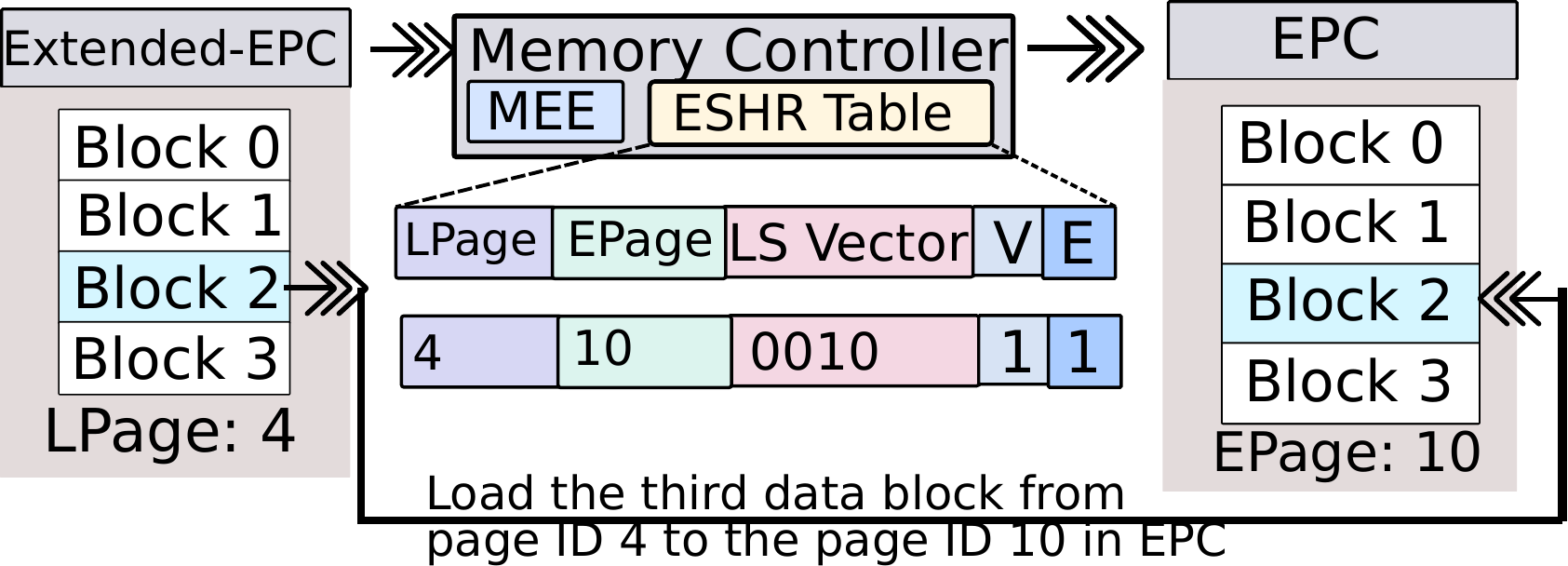}
    \caption{Handling EPC page faults using the ESHR table}
    \label{fig:eshr}
    \vspace{-2mm}
\end{figure}

The ESHR table stores
32 entries.  Each entry in the ESHR table contains five fields: \textit{EPage} 
represents the page ID of the evicted page; \textit{LPage} represents the page ID of the page that is being loaded
(newly mapped); \textit{LS Vector} is a 64-bit vector that indicates the load status of $LPage$; \textit{E-bit} is an
eviction bit indicating we need an eviction (of the {\em EPage}); and \textit{V-bit} is a valid bit indicating if the page is fully loaded or not.
We present a dummy example in Figure~\ref{fig:eshr}, where each page comprises four blocks. The page blocks in the
{\em LPage}, which are located in the eEPC are loaded to the EPC at the position pointed to
by the {\em EPage}.  

%Along with the page ID, the respective timestamps of the pages are also stored in the ESHR.
%FIXME: What is a timestamp. Never seen this before. Also, EPage does not have a consistent
%spelling. Somewhere it is Epage and somewhere EPage

\noindent \textbf{Execution Flow:} 
\subsubsection{Read Path}
When an EPC miss occurs for a read request, the data is fetched immediately from the eEPC region as it lies on the critical path and decrypted (Figure~\ref{fig:flowchart} shows the steps).

\subsubsection{Write Path}
A write request does not lie on the critical path. In case of an EPC miss for a write request, we
do the following:
\circlenew{1} If there are no other requests queued in the memory bus, the write request proceeds as usual. 
\circlenew{2} If a previous operation is still ongoing, an entry for the request is made in the ESHR; the request is made to wait til the process completes. 
\circlenew{3} If a read request arrives while this write is waiting, the read request is given higher priority . Additionally, if the two requests are for a memory region covered by the same subtree, they are grouped together such that for the higher level MACs of the subtree, a single verification operation would suffice for both requests.

\subsubsection{Verification Path:}
The MAC verification of the newly mapped page is carried out concurrently by the MVC, while the page blocks are being moved to the EPC.
The execution flow is not hindered by this process. Additionally, the remaining page blocks are loaded/evicted in parallel, while verification and MAC computation continues.
Thus, while fetching data from the eEPC region, the execution can be restarted
after just two memory reads (requested page block and its key). This drastically reduces
the latency of the critical path. As the overhead incurred during EPC page faults is comparable to the overhead of fetching data from the EPC itself, the overall impact of the large eEPC is very small.

\subsubsection{Communication with the OS}
We maintain a separate memory region with very few pages that
has relaxed security guarantees. 
This memory region is used for communication between trusted enclaves or between an 
enclave and the untrusted OS. The enclaves use this memory as a scratchpad for sending system call arguments
and receiving data from the OS and other enclaves. Similar to SGX-Client, this memory region can either have no security
or it could be encrypted with a session key.

\vspace{-2mm}
\subsection{Design Optimizations}
MAC verification does not lie on the critical path, but it still accounts for DRAM
accesses. These accesses can delay regular accesses. They also increase DRAM power consumption.
Hence, there is a need to minimize such additional accesses. 

\subsubsection{Optimizing MAC Verification}
We maintain a small cache in the TCB that stores {\em r} recently accessed top-level MACs of the MAC forest (i.e., the roots of the recently accessed subtrees in the MAC forest). 
Each MAC at the top level of the forest is the root MAC for a 512 KB memory region in the
eEPC region. This top-level MAC, which is stored in the EPC, needs to be retrieved from the main memory while performing MAC verification of any of the pages belonging to the region covered by this subtree root
(defined as its {\em subtree region}). Hence, we reduce one DRAM memory access by caching it.

\subsubsection{Optimizing MAC Forest Updates}
Updates to the MAC forest are required when pages are evicted from the EPC. 
The page to be evicted is selected based on an LRU
(least recently used) mechanism (stored in the EPC's metadata, EPCM). 
We additionally store the store the ID of the page that is next
in line for eviction in an {\em evict register} (computation is off the critical path). 
If both the currently evicted page and the next page to be evicted lie in the memory
region protected by the same subtree of the MAC forest, then the MAC updates to higher levels of the subtree for both the the pages can be clubbed together -- this reduces the number of memory accesses at the higher levels of the MAC forest.

\subsubsection{Corner Cases}
\circlenew{1} If speculative execution is in progress when a system call occurs, the processor first waits for the MAC verification to complete before taking any action. 

\circlenew{2} Without waiting to pair a write with an eviction (for reducing DRAM writes), we finish all verification operations as soon as possible.

\vspace{-3mm}
\section{Evaluation}
\label{sec:eval}

We compare the performance of \fname with SGX-Client as well as state-of-the-art
work (DFP and Penglai) by simulating these systems in our
cycle-approximate simulator Tejas\cite{tejas}. 
{\em Performance} is proportional to
the reciprocal of the simulated execution time. The system specifications used for evaluation are the same as that used during characterization (Table~\ref{tab:specs} from $\S$\ref{sec:char}).
The unsecure system is considered to be the {\em baseline}. 
\vspace{-2mm}
\subsection{Performance Analysis}
\begin{figure}[!ht]
    \centering
    \includegraphics[width=0.9\linewidth]{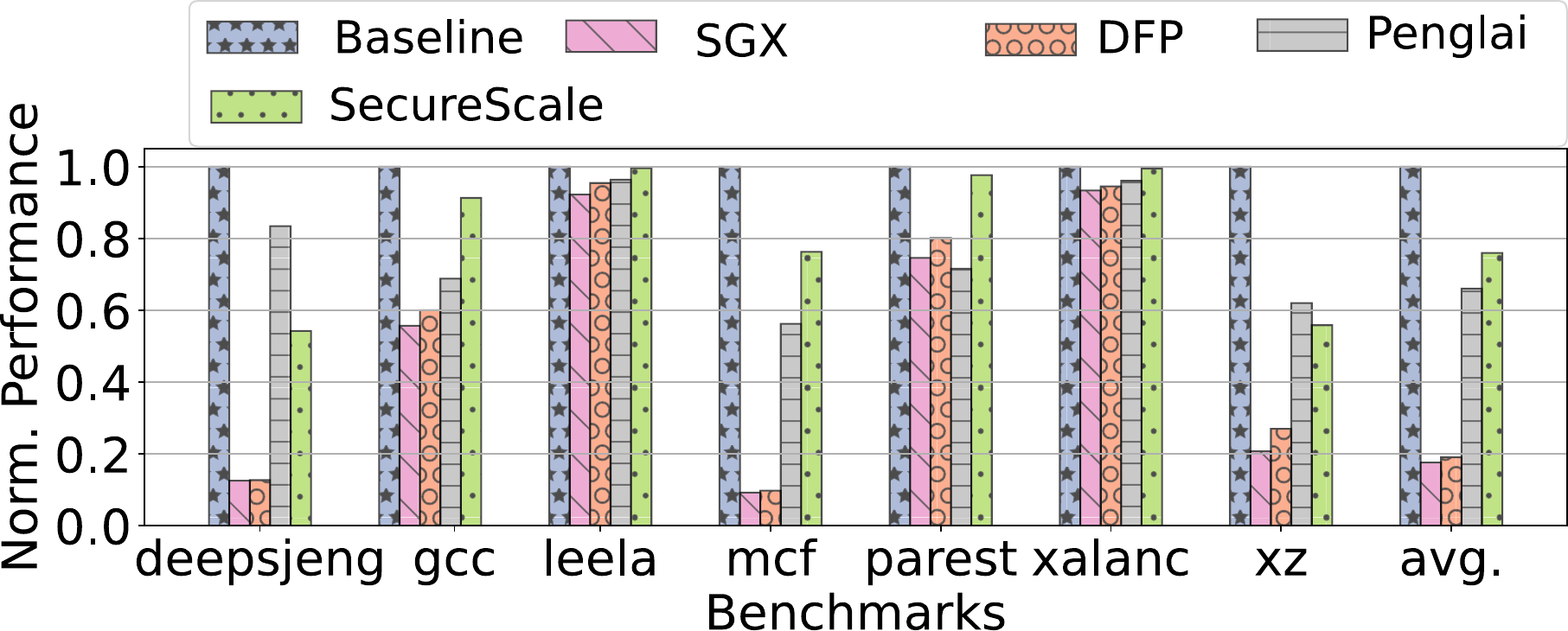}
    \caption{Performance of different systems. (\fname exhibits performance improvement over all the systems.)}
    \label{fig:ipc}
\end{figure}

In most workloads, the memory accesses are not very irregular. Hence,
the frequency of EPC page faults is low in general. The exceptions arise in the cases where the memory accessed is very large and the pattern of page accesses is random. These benchmarks experience an increased number of EPC page faults and incur a higher memory traffic overhead for integrity verification. 
Consequently, the benchmarks with a larger number of page faults will show a greater degradation in performance, as we can see in Figure~\ref{fig:ipc}. 
\vspace{-2mm}
SGX exhibits a drastic degradation in performance compared to the baseline (83\%).
Comparing the performance of the related work with that of SGX, we  see that DFP shows a
2$\%$ improvement in performance and Penglai performs 47$\%$ better than DFP. 
\fname shows improved performance compared to both DFP and Penglai (57\% and 10\%, resp.).
In 5/7 of the workloads, \fname performs much better than all the others
and exhibits the lowest degradation in average performance(24$\%$) with respect to the baseline. 
On an average, \fname \textit{exhibits a 59$\%$ improvement in performance}
vis-a-vis SGX-Client.

%FIXME: Take the names of a few benchmarks. Give benchmark-specific numbers/analysis.
\vspace{-2mm}
\subsection{Detailed Analysis}

\subsubsection{Performance Degradation}
\begin{figure}[!htb]
    \centering
    \includegraphics[width=0.7\linewidth]{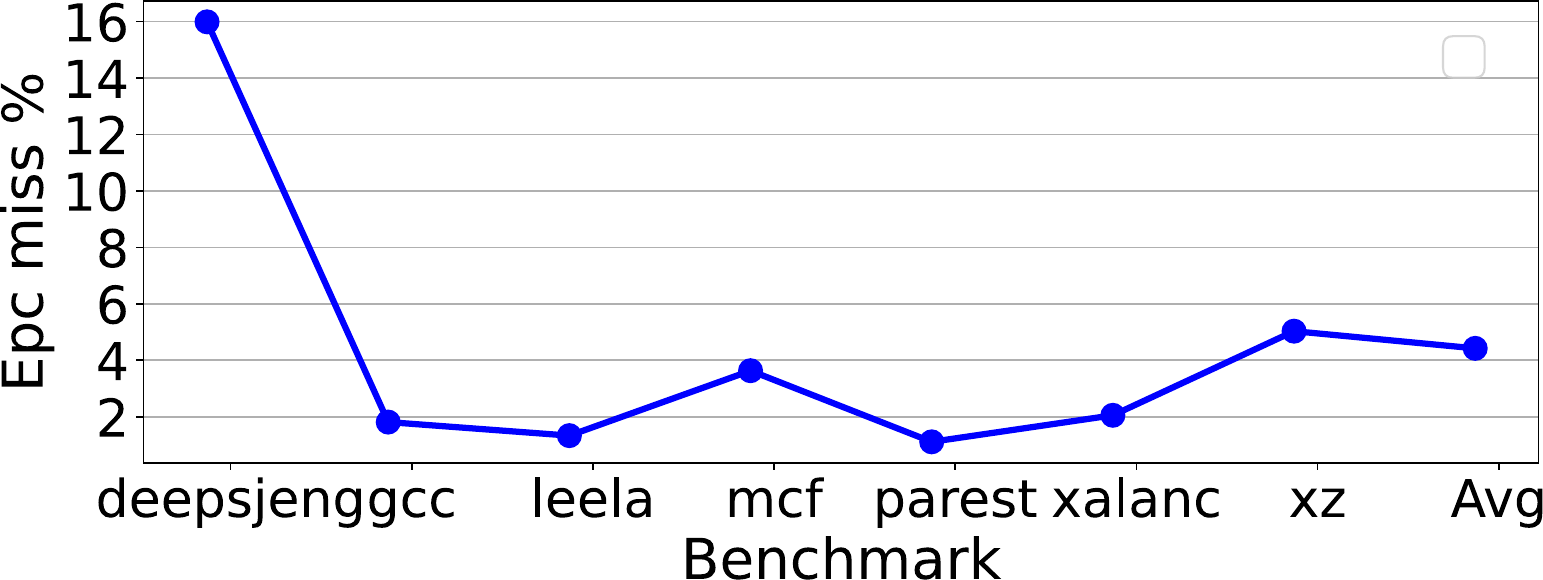}
    \caption{EPC miss $\%$ w.r.t. LLC misses for different benchmarks. (The benchmarks with {\em higher} EPC miss rates {\em(page faults)} are the ones with a {\em greater performance degradation}.)} 
    \label{fig:EpcLlc}
    \vspace{-2mm}
\end{figure}

The EPC page fault rate, in terms of EPC misses $\%$ with respect to LLC misses is shown in Figure \ref{fig:EpcLlc}. 
This map gives us an estimate of the spatial locality in the benchmark suite.
The benchmarks that access more memory pages are associated with
a higher EPC page fault $\%$. The access pattern (randomness) of the pages also affects the EPC page faults. 
The benchmarks with a high percentage of EPC misses with respect to LLC misses like {\em deepsjeng} experience drastic degradation in performance (88$\%$ degradation w.r.t. baseline in SGX) as shown in Figure \ref{fig:ipc}. On the other hand {\em leela}, which has a very low EPC miss rate w.r.t. to LLC misses, experiences a very low degradation (8$\%$ degradation w.r.t. baseline in SGX).

\subsubsection{Optimizations}

\begin{figure}[!htb]
    \centering
    \includegraphics[width=0.9\linewidth]{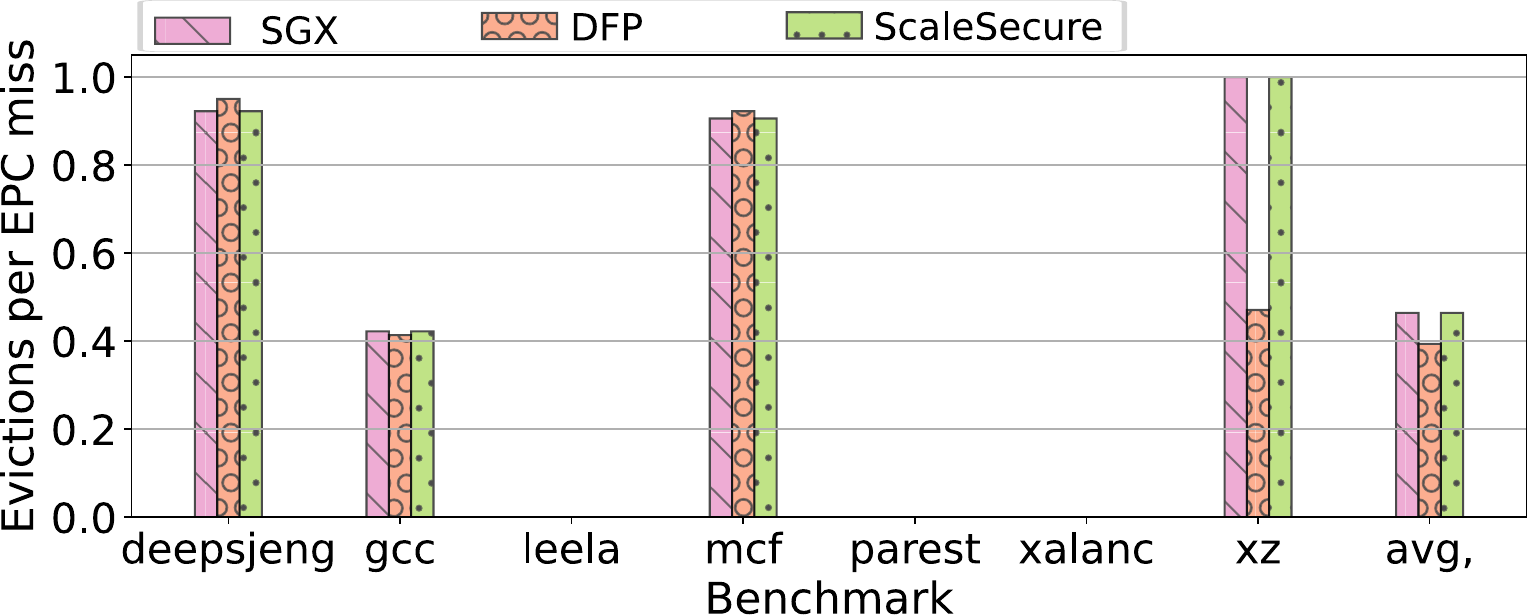}
    \caption{Evictions per EPC miss. {\em (Eviction rate directly affects the extent of performance degradation.)}}
    \label{fig:evictR}
\end{figure}

Figure~\ref{fig:evictR} shows the evictions per EPC miss in the various benchmarks for different models. 
In systems with an EPC, the eviction rate directly affects system performance. 
\vspace{-4mm}

Note that the number of evictions could be less than 1, if we have already created space for the page using prefetching (like in DFP).
Specifically, the eviction rate varies (increases/decreases) in DFP, in comparison to SGX, because of its predictions/mis-predictions. 
In {\em deepjeng and mcf}, the evictions increase for DFP because of mispredictions, whereas in {\em xz}, asynchrounous preloading of correctly predicted faulting pages reduces the evictions in 
the critical path. DFP thus shows better performance than SGX in case of {\em xz} even though both SGX and DFP impose the same penalty for evictions.
However, in \fname, the system performs much better than both SGX and DFP in all the benchmarks (even though it has the same eviction rate as SGX), including {\em deepsjeng(42$\%$), mcf(65$\%$) and xz(29$\%$)}, because it drastically reduces the penalty associated with EPC misses and evictions. 

The page access pattern also affects the Merkle tree access overheads in both SGX and DFP. In Penglai, it impacts the MMT access overheads. The additional bandwidth associated with integrity verification varies depending on this pattern as can be seen in the case of {\em parest} where it performs worse than SGX by 3$\%$ and DFP by 9$\%$ even though it does not have an EPC (and EPC associated overheads). 
In case of \fname, the Merkle tree protects the counters of only those pages that reside inside the EPC and therefore it has a fixed size. {\em This ensures that the Merkle tree overheads are minimized in} \fname. 
The result of these optimizations is evident in the performance improvement seen in {\em parest} for \fname (23$\%$ over SGX, 17$\%$ over DFP and 26$\%$ over Penglai).
\vspace{-2mm}
\subsection{Optimizations for Reducing MAC Accesses}
The MACs are accessed during the MAC verification and MAC update phases. These processes require fetching MACs from different levels of the MAC forest (stored in memory). Although these
accesses are not on the critical path, every memory access increases DRAM traffic and DRAM power. We introduced optimizations in the design to reduce the number of additional memory accesses required to perform these operations.

\subsubsection{Optimizing MAC Updates}
\begin{figure}[!htb]
    \centering
    \includegraphics[width=0.8\linewidth]{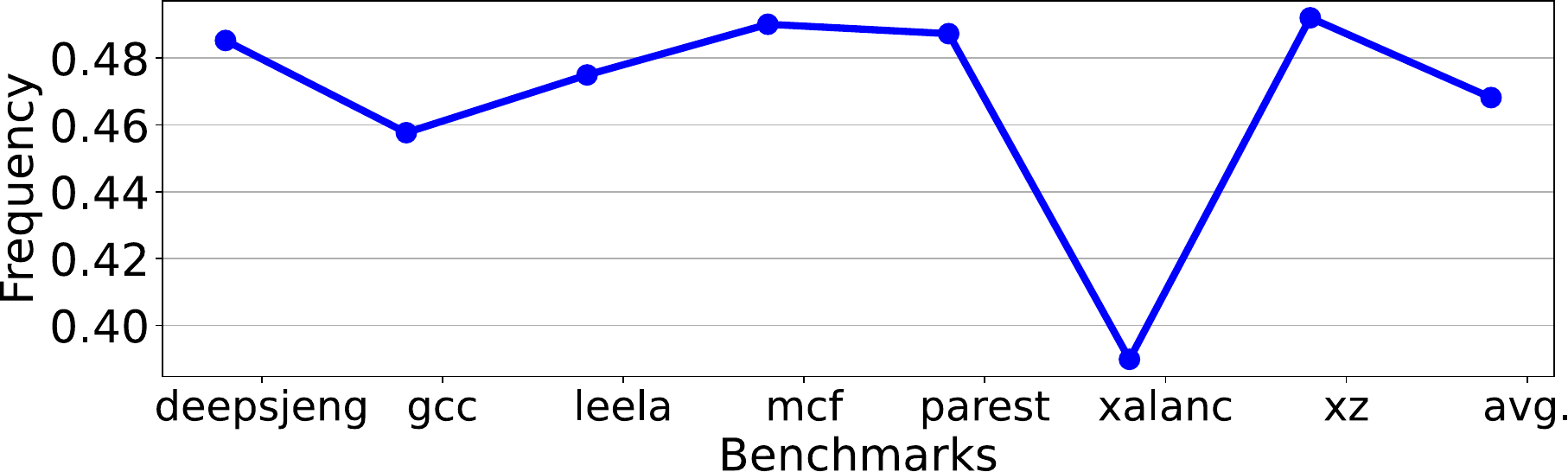}
    \caption{Frequency of clubbing {\em (clubbing of updates is possible for the pages protected by the same subtree.)}}
    \label{fig:club_freq}
\end{figure}

We club the MAC updates for the higher levels of the subtrees in the MAC forest during consecutive evictions of  pages that belong to the same {\em subtree region}. This reduces the number of memory accesses required for updating the MACs in the higher levels of the MAC forest. Figure~\ref{fig:club_freq} shows the frequency of clubbing of the updates observed in our workloads (an average of 46$\%$ clubbing was observed). {\em This plot depicts the spatial locality in the benchmarks within a subtree region (512 KB memory region).} Note that clubbing of updates is possible only if consecutive pages fall in this region. 

\subsubsection{Optimizing MAC Verification}
\begin{figure}[!htb]
    \centering
    \includegraphics[width=0.8\linewidth]{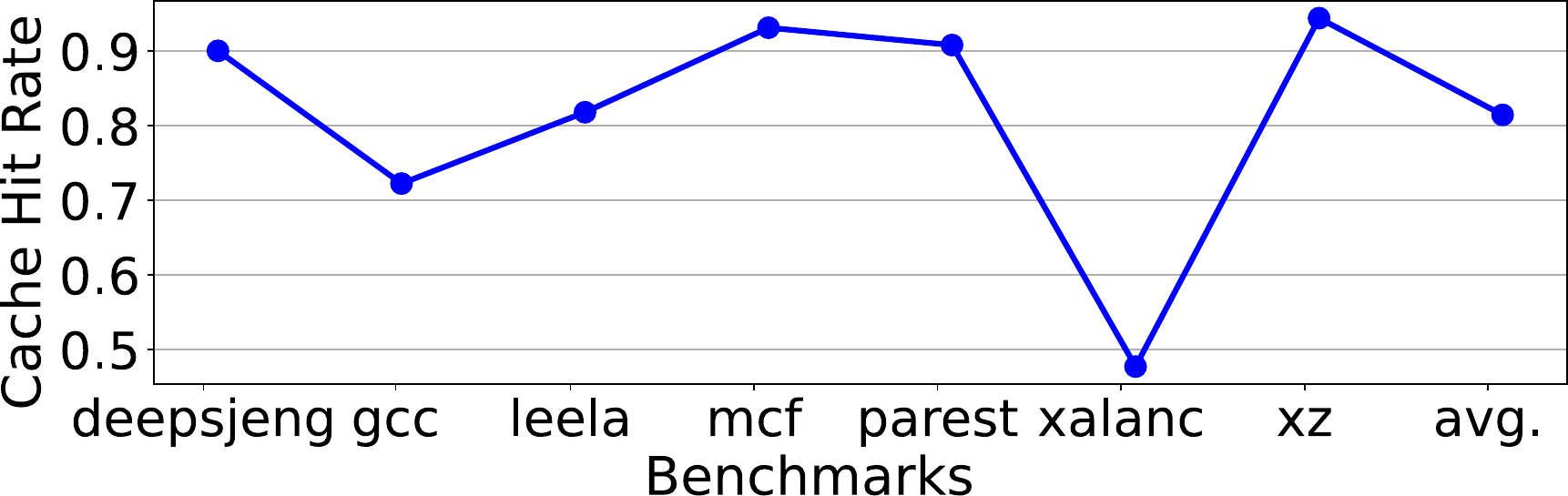}
    \caption{Top-level MAC cache hit rate}
    \label{fig:hit_rate}
\end{figure}

We introduced a small cache in the TCB to store the 8 recently accessed MACs from the top-most level of the MAC forest in the TCB. We attempt to leverage any locality of accesses that might exist in the workloads for the pages secured by the subtrees of the cached top-level nodes. 
This reduces the number of memory accesses required to retrieve the top-level nodes from the EPC for MAC verification. Figure~\ref{fig:hit_rate} shows the cache hit rates in our design for various workloads. The average hit rate is 81.4$\%$. 

\begin{figure}[!htb]
    \centering
    \includegraphics[width=0.8\linewidth]{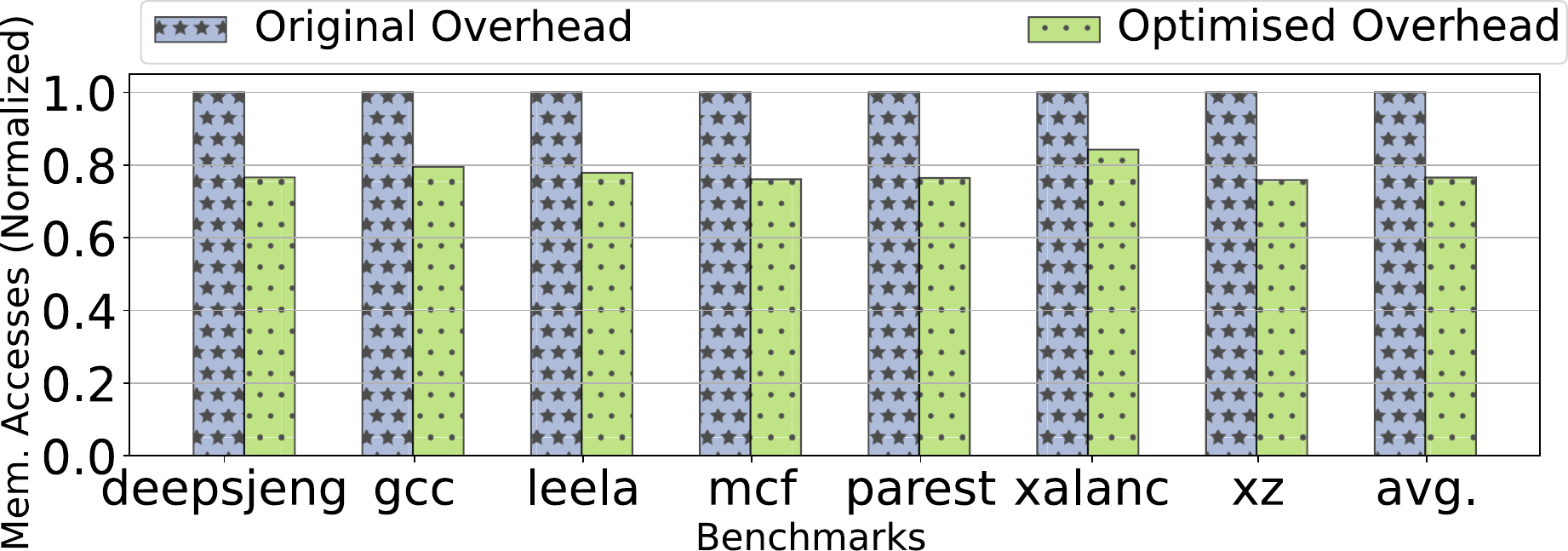}
    \caption{Reduction in DRAM traffic due to optimizations for MAC accesses. {\em (Result: additional accesses reduced by 23.5$\%$ in our workloads.)} }
    \vspace{-5mm}
    \label{fig:dram}
\end{figure}

These optimizations leverage the temporal and spatial locality of EPC misses in order to reduce the number of memory accesses to the higher level nodes in the MAC forest. 
If we observe closely, both these figures show a similar pattern -- they depict the extent of locality in each workload.
Comparing the pattern with DRAM traffic reduction (shown in Figure~\ref{fig:dram}), we see that the benchmark {\em mcf} that has a high MAC cache hit rate and frequency of clubbing exhibits a larger decrease in DRAM traffic (24$\%$), compared to xalanx (16$\%$) which has a low cache hit rate and clubbing frequency. 
We observe that by employing both these optimizations, the overall additional memory accesses to retrieve the MACs reduces by 23.5$\%$ in our workloads.
\vspace{-2mm}
\subsection{Sensitivity Analysis}

The hyperparameters in our system comprise {\em the number of levels} and {\em the arity} of the subtrees in the MAC forest, which are used during the integrity verification of the pages. We compute an 8-byte MAC for every 4 KB page. A 512 GB memory, contains $2^ {27}$ MACs (1 MAC per page), which is the number of nodes in the lowest level of the MAC forest. 
We set the hyperparameters -- {\em q} and {\em p} -- such that the performance overhead is minimized.

\noindent{\em \textbf{Subtree Level Analysis (q)}} 
As we have observed in SGX and Penglai, the number of levels in the integrity trees/forests correlate very well with the memory access overheads. 
Additionally, the number of levels also influence the storage overheads. 
Keeping both in mind, we chose {\em \textbf{q}} as \textbf{3} because it maximized our performance.

\begin{figure}[!htb]
    \centering
    \includegraphics[width=0.7\linewidth]{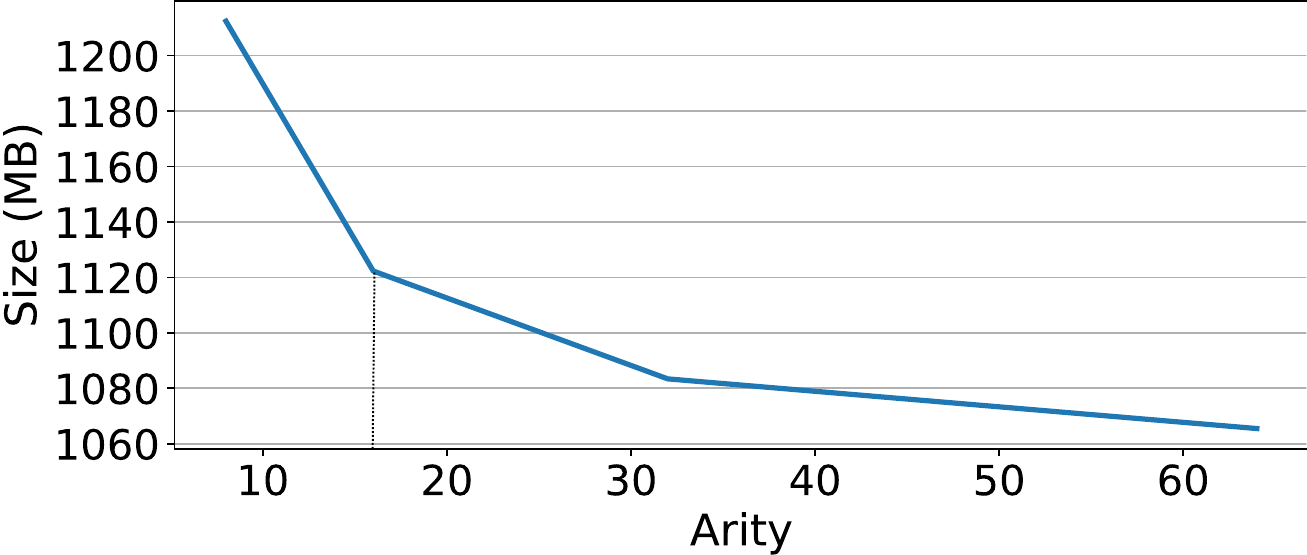}
    \caption{Storage overhead of varying arity of subtrees in the MAC forest}
    \label{fig:macstore}
    \vspace{-3mm}
\end{figure}

\noindent{\em \textbf{Subtree Arity Analysis (p)}} 
The storage and maintenance of the MACs is another concern for the MAC forest. 
The arity of the subtrees also dictates the number of memory accesses required while verifying or updating the MACs in the subtree region.
Thus, we decide to keep the arity small.
We set the arity of the higher level as half of that of the lower level to reduce the frequency of updates in the higher level nodes. Additionally, we plotted the
the storage overhead for different values of  the hyperparameter {\em p} (see Figure~\ref{fig:macstore}).   
The storage overhead shows a sharp decline in the beginning after which the descent is more gradual. We thus decided to select the arity close to the knee of the curve and set it to 16 for the lower level
and 8 for the level above it.. 

The total storage space required for our forest is 1096 MB (for all three levels of the tree). Our Merkle Tree (for the EPC) has arity ($32 \times 32 \times 32$) and a size of 2.06 MB. Thus the combined storage overhead of both these structures for 512 GB memory in our design sums up to 1098.06 MB.
This is 8 times smaller than what the SGX-Client Merkle Tree would require (8322.06 MB) for securing 512 GB memory.
\vspace{-1.8 mm}
\subsection{Security Analysis}
%\fname provide all the ACIF guarantees --
\fname ensures robust security guarantees (ACIF) across the complete system. Note that the EPC provides all four guarantees because it uses the same system as SGX-Client. Let us thus focus on the eEPC.

\noindent $\blacktriangleright$  \textbf{Authenticity (A)} For authentication of the enclave pages in the eEPC, the key contains a HW-specific key (device-level), a boot-specific component used to seed the PRNG, and an enclave specific enclave ID (enclave-level). This ensures that only the enclave that owns the page can access it. The MAC check for any other enclave including the OS will fail. They will not have a valid enclave id to construct the key that is needed to access (read/write) the page and recompute the MAC for verification. Thus, we cryptographically ensure authenticity. \\
\noindent $\blacktriangleright$ {\bf Confidentiality (C)} is guaranteed by encrypting the data using  standard AES-CTR mode encryption for the EPC and AES-ECB mode encryption for eEPC regions. This ensures that only the writing enclave can decrypt and access the original plaintext data. \\
\noindent $\blacktriangleright$ \textbf{Integrity (I)} In order to protect the data integrity of the eEPC, we maintain 
page-level MACs for each eEPC page. These MACs are protected with a multi-level MAC forest whose top-level nodes
are stored in the EPC. Hence, any integrity violation will be caught in the MAC verification phase. The integrity of the Key Table (stored in the EPC) is established using the key to encrypt the hash of the page and construct the lowest level MAC. \\
\noindent $\blacktriangleright$ {\bf Freshness (F)} is guaranteed by generating  
a new key every time a page is written back to the eEPC. We use a PRNG to generate the new key along with a bunch of other fields. We consider this to be secure enough given that we don't expect the same key to repeat in any practically relevant duration of time. However, if more security is desired then a global counter can be used.

\begin{scriptsize}
\begin{table*}[!htb]
%\footnotesize
    \centering
    \setlength\tabcolsep{1.5pt} 
\begin{tcolorbox}[enhanced, width=0.732\textwidth, boxsep=0pt, left=0pt,right=0pt,top=0pt,bottom=0pt,
    colback=white, colframe=black, arc=0mm, boxrule=0pt,
    drop shadow={shadow xshift=140mm, shadow yshift=100mm}]
    %\footnotesize
 \rowcolors{2}{royalblue!20}{royalblue!5}

    \begin{tabular}{|l|c|c|c|c|c|c|c|c|c|}
    \hline
     \rowcolor{gray!10}
         \multicolumn{2}{|c|}{}  & \multicolumn{3}{c|}{{\em \textbf{Security Components}}}& \multicolumn{2}{c|}{{\em \textbf{Security Characteristics}}} & \multicolumn{2}{c|}{{\em \textbf{Enclave Enhancement}}} \\

    \hline
    \rowcolor{gray!10}
         \textbf{System} & \textbf{Arch} &\textbf{Confidentiality}  &\textbf{Integrity} &\textbf{Freshness} & \textbf{HW} & \textbf{Standalone} & \textbf{Scalable} &  \textbf{EPC Page fault} \\ 
          \rowcolor{gray!10}
          & & & & &\textbf{(exclusively)} & \textbf{Security} & \textbf{enclave} & \textbf{latency redn.}\\
         \hline           
                      
                     \hline
         SGX-Client\cite{sgx} & Intel & \textcolor{teal}{$\checkmark$} & \textcolor{teal}{$\checkmark$} &  \textcolor{teal}{$\checkmark$}  &\textcolor{teal}{$\checkmark$} & \textcolor{teal}{$\checkmark$} &  \textcolor{red}{$\times$} & \textcolor{red}{$\times$} \\

          \hline       
         SGX-Server\cite{sgx1} & Intel & \textcolor{teal}{$\checkmark$} & \textcolor{teal}{$\checkmark$} &  \textcolor{red}{$\times$} & \textcolor{teal}{$\checkmark$}  & \textcolor{teal}{$\checkmark$} & \textcolor{teal}{$\checkmark$}  & \textcolor{red}{$\times$} \\

         \hline       
         TDX\cite{tdx1} & Intel & \textcolor{teal}{$\checkmark$} & \textcolor{teal}{$\checkmark$} &  \textcolor{red}{$\times$} & \textcolor{red}{$\times$} & \textcolor{teal}{$\checkmark$} & \textcolor{teal}{$\checkmark$}  & \textcolor{red}{$\times$} \\

         \hline       
         SEV-SNP\cite{snp} & AMD & \textcolor{teal}{$\checkmark$} & \textcolor{teal}{$\checkmark$} &  \textcolor{red}{$\times$} & \textcolor{red}{$\times$} & \textcolor{teal}{$\checkmark$} & \textcolor{teal}{$\checkmark$}  & - \\
        
          \hline
         CoSMIX\cite{cosmix} & Intel & \textcolor{teal}{$\checkmark$} & \textcolor{teal}{$\checkmark$} &  \textcolor{teal}{$\checkmark$} & \textcolor{red}{$\times$} & \textcolor{teal}{$\checkmark$} & \textcolor{red}{$\times$}  & \textcolor{red}{$\times$} \\

       %  \hline
       %  CosMIX\cite{cosmix} & Intel & \textcolor{teal}{$\checkmark$} & \textcolor{teal}{$\checkmark$} &  \textcolor{teal}{$\checkmark$} & \textcolor{red}{$\times$} & \textcolor{teal}{$\checkmark$} & \textcolor{red}{$\times$}  & \textcolor{red}{$\times$} \\
         
     %    \hline
     %    ShieldStore\cite{shieldstore} & Intel & \textcolor{teal}{$\checkmark$} & \textcolor{teal}{$\checkmark$} &  \textcolor{teal}{$\checkmark$} & \textcolor{red}{$\times$} & \textcolor{teal}{$\checkmark$} & \textcolor{red}{$\times$}  & \textcolor{red}{$\times$} \\

       %  \hline
       %  Sanctuary\cite{sanctuary} & ARM & \textcolor{teal}{$\checkmark$} & \textcolor{red}{$\times$} &  \textcolor{red}{$\times$} & \textcolor{red}{$\times$} & \textcolor{teal}{$\checkmark$} & \textcolor{red}{$\times$}  & - \\
         
         \hline     
         DFP\cite{dfp} & Intel & \textcolor{teal}{$\checkmark$} & \textcolor{teal}{$\checkmark$} &  \textcolor{teal}{$\checkmark$} & \textcolor{red}{$\times$} & \textcolor{teal}{$\checkmark$} & \textcolor{red}{$\times$} &  \textcolor{red}{$\times$} \\

           \hline
         Penglai\cite{penglai} & RISC-V & \textcolor{teal}{$\checkmark$} & \textcolor{teal}{$\checkmark$} & \textcolor{teal}{$\checkmark$} & \textcolor{red}{$\times$} & \textcolor{red}{$\times$} & \textcolor{teal}{$\checkmark$} &  - \\ 

         \hline     
         ARM-CCA\cite{arm_cca} & ARM & \textcolor{teal}{$\checkmark$} & \textcolor{teal}{$\checkmark$} &  \textcolor{red}{$\times$} & \textcolor{red}{$\times$} & \textcolor{teal}{$\checkmark$} & \textcolor{teal}{$\checkmark$} &  - \\

          \hline
         \rowcolor{gray!20}
         \textbf{\fname} & Intel & \textcolor{teal}{$\checkmark$} & \textcolor{teal}{$\checkmark$} &  \textcolor{teal}{$\checkmark$} &\textcolor{teal}{$\checkmark$} & \textcolor{teal}{$\checkmark$} & \textcolor{teal}{$\checkmark$} & \textcolor{teal}{$\checkmark$} \\
         
    \hline
    \end{tabular}
    \end{tcolorbox}
    \caption{Comparison of related work. `-' means not applicable}
    \label{table:compare}
    \vskip -4mm
\end{table*}
\end{scriptsize}

\noindent \textbf{Security Analysis of the Page Table} --
The page table needs to be protected in designs that have large unrestricted unsecure memories or in cases where enclave isolation is not guaranteed (e.g. Intel SGX and Penglai). We argue that in \fname, we do not need this kind of protection because of the following reasons. Here are the possible attacks that the OS can mount through the page table.

\noindent \circlenew{A} {\em Secure $\rightarrow$ Unsecure Mapping:}  This is not relevant in our case because in our system
the entire memory is protected. However, this is a genuine problem in systems like Intel SGX and ARM TrustZone because
they have large unsecure memories.

\noindent \circlenew{B} {\em Unsecure $\rightarrow$ Secure Mapping:} This cannot happen for the same reason
outlined in the previous point.

\noindent \circlenew{C} {\em Secure $\rightarrow$ Secure Mapping:} Another possibility is when the OS maps the secure page of an enclave to the secure region of another enclave. The unauthorized enclave
cannot read or write the contents of the page since the enclave ID is a part of the key. The MAC check will fail
and this will be a catastrophic event. An OS can only create an enclave or fully tear it down -- it cannot access any page within it. For maintenance of enclave IDs, we use the same system as SGX-Client.

%\circlenew{1} Confidentiality (C) is preserved within both the EPC and the eEPC region through the utilization of standard AES-CTR mode encryption for EPC and AES-ECB mode encryption for eEPC. The counters in the AES-CTR mode are protected using the Merkle tree while the encryption keys used in the AES-ECB mode are encrypted using a special session-specific key. This ensures that only a verified user can decrypt and access the original plaintext data.%
%\circlenew{2} {\em Integrity (I)} is ensured by employing MAC verification using MVC and the MAC forest.
 %\circlenew{3} {\em Freshness (F)} is guaranteed differently in EPC and eEPC regions. The \textbf{EPC} region is encrypted using AES-CTR, the counters provide freshness to the keys. The integrity of the counters is ensured using a Merkle tree. 
%A new key is generated for every encryption of every page stored in the \textbf{eEPC} region (AES-ECB). A PRNG is used in the key generation process. The keys are protected in the eEPC using encryption and MACs. 
%\circlenew{4} For {\em Authentication (A)} of the enclave pages in the eEPC, the key contains a HW-specific key, a boot-specific component used to seed the PRNG, and an enclave specific enclave ID. 

\vspace{-3mm}

\section{Related Work}
\label{sec:RW}

The size of the enclaves can be enhanced using two main approaches - by using bespoke secure systems designed for server applications or by using certain optimization techniques to enhance the enclave size
(refer to Table~\ref{table:compare}).
\vspace{-2mm}
\subsection{Bespoke Systems}
Most state-of-the-art secure servers focus on virtual machine (VM) isolation.
The TCB also includes the guest VM along with its software stack.

AMD's {\em SEV-SNP (Secure Encrypted Virtualization - Secure Nested Paging)} \cite{snp} supports both main memory encryption and encrypted virtual machines (VMs). It does not provide freshness 
or protection against some physical attacks --
attacking the DDR bus while the VM is actively running.
Intel {\em TDX (Trust Domain Extensions)}\cite{tdx1} is similar and is vulnerable to replay
attacks.

ARM's recent Confidential Compute Architecture (ARM CCA)\cite{arm_cca} is also based on similar secure virtualization technologies. It introduces {\em Realms}, which enables isolated memory for secure execution, and a page-locking mechanism to support large enclaves (realms).
However, CCA does not employ encryption and cannot defend the system against physical attacks like cold boot attacks, live probing or replay attacks. 

A different approach is adopted in 
{\em Penglai} \cite{penglai}, which is a software-hardware co-designed  system that creates
dedicated hardware augmentations on a RISC-V core.
There is one large EPC -- protected by
counters and a single Merkle tree. Its recipe for scalability is to mount sub-trees of the Merkle
tree on demand -- these are called mountable Merkle trees (MMTs). It furthermore caches a few
MMT roots in the TCB.
In the event of an LLC miss, if the MMT root is found in the MMT cache, then the counters can be verified with additional memory accesses. However, if there is a miss, then the penalty is quite large.
Note that all of this is on the critical path. Hence, the read latencies are quite high (something 
that we have seen in our experiments as well). Additionally, it relies on dedicated HW support to ensure
that the memory region that stores the MMTs is not tampered with. This is not possible in our threat model where we allow the attacker to modify any memory location at will.

Summary: These systems either do not provide all four ACIF guarantees or are not compliant with our threat model.
\vspace{-2mm}
\subsection{Enclave Size Enhancement via Memory System Optimizations}

{\em CoSMIX} \cite{cosmix} proposes a software cache to store evicted EPC pages. 
However, providing 
the same level of security in software as provided by hardware is seldom possible~\cite{dfp}.
Hence, \fname solely relies on hardware solutions.
Liu et al.~\cite{dfp} (DFP) attempt to decrease the number of EPC page faults on the critical path
by prefetching pages into the EPC. They leverage sequential access patterns and use a list-based prefetcher. We have compared our work with DFP in Section~\ref{sec:eval} and shown large
performance gains mainly because the accuracy of the predictor is low.

\vspace{-3mm}
\section{Conclusion}
\label{sec:concl}

We introduced three new ideas in this paper, which allowed us to solve a problem that was known for
a long time but had become a matter of great concern ever since Intel deprecated SGX-Client in 2021. Sacrificing {\em freshness} is the industry standard as of today mainly because providing it requires
maintaining counters for every block and a Merkle tree, which are not scalable by design. We 
leveraged the fact that the catastrophic nature of a security verification failure can be used to 
do a {\em little bit more speculation} and take verification totally off the critical path. Second, the state-of-the-art has put its full might behind protecting the integrity of aspects of the key such as the counters. However, we opt for a diametrically different approach, where a read arrives to the processor with a key that has {\em supposedly been used to encrypt it}. This allowed us to create a MAC forest where we could verify the integrity of the key and the data together in
a delayed fashion. These three ideas along with some design optimizations to reduce DRAM accesses allowed us to achieve a 10\% speedup over our nearest competitor Penglai and a 59\% speedup over a vanilla SGX-Client implementation.

%\newpage

%%%%%%% -- PAPER CONTENT ENDS -- %%%%%%%%

%%%%%%%%% -- BIB STYLE AND FILE -- %%%%%%%%
\bibliographystyle{IEEEtranS}
\bibliography{refs}
%%%%%%%%%%%%%%%%%%%%%%%%%%%%%%%%%%%%

\end{document}